\documentclass[11pt,a4paper]{article}
\usepackage[margin=3.0cm]{geometry}
\usepackage[T1]{fontenc}
\usepackage[utf8]{inputenc}
\usepackage{lmodern}
\usepackage{amsmath,amssymb}
\usepackage{graphicx}
\usepackage{booktabs}
\usepackage{array}
\usepackage{longtable}
\usepackage{caption}
\usepackage{xcolor}
\usepackage{url}
\usepackage[hidelinks]{hyperref}

\newcommand{\eps}{\varepsilon}
\newcommand{\BQ}{B_{Q}}
\newcommand{\dlog}[1]{\mathrm{d}\log(#1)}
\newcommand{\dlogn}{\mathrm{d}\log n}
\newcommand{\ci}[2]{(#1\text{--}#2)}
\newcommand{\cito}[2]{(#1\text{ to }#2)}
\newcommand{\device}[1]{\texttt{#1}}
\newcommand{\rec}[1]{\path{#1}}

\newcommand{\SIfailures}{Supplementary Note~1}
\newcommand{\SIfixed}{Supplementary Note~2}
\newcommand{\SIeffdim}{Supplementary Note~3}
\newcommand{\SItrained}{Supplementary Note~4}
\newcommand{\SInotation}{Supplementary Note~5}
\newcommand{\SIclaims}{Supplementary Table~S4}
\newcommand{\PaperTitle}{Shot noise can decide capacity comparisons of quantum models}

\newcommand{\MainAuthors}{%
  Pavel Sulimov$^{1}$\thanks{Corresponding author. Email: suli@zhaw.ch.
    ORCID 0000-0003-2885-2646.}%
  \and Claude Lehmann$^{1,2}$\thanks{ORCID 0000-0002-4693-0444.}%
}
\newcommand{\AffiliationBlock}{%
  $^{1}$ Zurich University of Applied Sciences (ZHAW), Winterthur, Switzerland\\
  $^{2}$ University of Zurich, Zurich, Switzerland}
\newcommand{\VersionNote}{}
\newcommand{\runin}[1]{\par\medskip\noindent\textbf{#1}\enspace\ignorespaces}
\newcommand{\keybox}[1]{\noindent\fbox{\begin{minipage}{\dimexpr\linewidth-2\fboxsep-2\fboxrule\relax}\small #1\end{minipage}}}

\newenvironment{regtable}{\begin{tabular}{@{}>{\raggedright\arraybackslash}p{5.2cm}>{\raggedright\arraybackslash}p{2.4cm}>{\raggedright\arraybackslash}p{6.4cm}@{}}}{\end{tabular}}

\title{\bfseries \PaperTitle}
\author{\MainAuthors}
\date{\AffiliationBlock}

\renewcommand{\SIfailures}{Appendix~\ref{app:failures}}
\renewcommand{\SIfixed}{Appendix~\ref{app:fixed}}
\renewcommand{\SIeffdim}{Appendix~\ref{app:effdim}}
\renewcommand{\SItrained}{Appendix~\ref{app:trained}}
\renewcommand{\SInotation}{Appendix~\ref{app:notation}}
\renewcommand{\SIclaims}{Table~\ref{tab:claims}}
\renewcommand{\VersionNote}{This version supersedes version~1 of this preprint
(arXiv:2609.14424v1), which combined the present results with those of the
companion paper and led with the cost of estimating the Fisher matrix. That cost
law is Theorem~1 here, with its sweeps in Appendix~\ref{app:fixed}; the
light-cone estimator and its hardware tests are in the companion paper.}

\begin{document}
\maketitle

\begin{abstract}
\noindent
Shot noise can decide close comparisons of what quantum models appear able to
learn. On an IBM processor, the standard estimate of a capacity measure, the
effective dimension, placed a 48-qubit model above about $90\%$ of classical
networks with as many parameters at 256 shots per circuit. The split-shot estimate
from the same data, which removes the bias of the Fisher matrix and needs no noise
model, placed it above $12\%$. A registered run at six parameter points of a
16-qubit model repeated this: the standard estimate placed the three random points
above every classical network, the split-shot estimate above $11$ to $22\%$. In
simulation the standard estimate reversed the comparison with the classical median
at all 100 random points tried. The inflation is the size that shot noise predicts, and a forecast
from a pilot run bounds it from above. For squared loss and a linear readout, the
error of the standard parameter-shift estimate of a model's Fisher matrix is an
exact polynomial in single-shot statistics. Its leading term grows with the
effective rank of the matrix, so the shot budget for a stated precision follows
from a pilot run; on hardware the observed errors were $0.85$ to $1.09$ times the
predicted ones. The effective dimension can be bracketed from the data but not
pinned down at a practical budget. Capacity reports should state their shot count
and use the split-shot estimate.

\end{abstract}

\section*{Introduction}

A variational quantum model is a quantum circuit whose gate angles are tuned on
data. How much such a model can express is commonly judged from one matrix, the
empirical Fisher matrix: the average over training samples of the outer product of
each sample's loss gradient with itself. The effective
dimension\cite{abbas2021,abbas2021b}, the most used capacity measure of this kind,
counts the directions in which that matrix stands above a resolution set by the
amount of data, and it has been used to compare quantum models with classical
networks of the same size\cite{abbas2021,mckiernan2026}. It is formally defined
from the Fisher information of the model's own predictive
distribution\cite{kunstner2019,meyer2021}. Both matrices are averages of outer
products of gradients measured by parameter shift, so the bias described below acts
on both; the exact error formula of Theorem~1 is stated for the empirical matrix
under squared loss with a linear readout. On the largest model run here the two
give the same effective dimension to within $0.1\%$, which is expected there rather
than a sign of generality: the model's outputs are close to zero, so both matrices
weight every sample alike. On a quantum computer each gradient entry is
estimated from repeated runs of a circuit, each ending in one measurement of the
qubits (a shot), and the number of shots behind a reported Fisher matrix is rarely
stated. The omission can decide a close comparison. On an IBM processor, a
48-qubit model whose capacity was estimated in the standard way from 256 shots per
circuit ranked above about $90\%$ of classical networks with as many parameters;
estimated from the same data with the split-shot estimate, which removes the bias
of the matrix without any noise model, it ranked above $12\%$
(Fig.~\ref{fig:decision}). That comparison was computed after the run, so we
registered a second one before running it, at six parameter points of a 16-qubit
model on another processor. There the standard estimate placed the three random
points above every classical network and the split-shot estimate above $11$ to
$22\%$. A
comparison flips only when the true gap between the models is smaller than the shot
inflation, and checking for this costs nothing, since the inflation is the
difference between the standard and the split-shot readings of the same data.

The reason is that shot noise does not average away in a nonlinear function of the
matrix. The standard estimate averages outer products of noisy gradients, so the
squared noise adds a positive term to every direction of the matrix. It lifts the
small eigenvalues, which are the ones a capacity measure counts most finely, and
the model looks more expressive than it is. On a trained 32-parameter model the
exact local effective dimension is $14.04$; from 256 shots per circuit it reads
$19.82$, $41\%$ high, and it comes within $0.5$ of the true value only at 16,384
shots per circuit (Fig.~\ref{fig:fisher}e). How far a capacity measure predicts
generalisation is itself debated\cite{gilfuster2024,mia2026}; whichever way that
debate goes, a measure that moves by tens of percent with the shot count cannot
settle it.

This paper measures the effect on hardware, gives the shot budget that prevents it,
and shows how to remove most of it when the budget is not met. The mechanism is
elementary: the mean of a squared estimated gradient exceeds the squared gradient
by its variance, and products of independent halves remove that excess, as in any
U-statistic\cite{hoeffding1948}. What is, to our knowledge, new is what follows
from it for this object. First, in two runs on IBM processors whose predictions
were fixed in advance, the standard estimate at 256 shots per circuit inflated the
effective dimension by $37\%$ and $46\%$, the size that shot noise alone predicts,
and on the larger model the inflation reversed a close capacity comparison. A
registered run at six parameter points, with every derivative read on all qubits,
and a simulation at 116 points show that at random parameter points the reversal
is the rule.
Second, for squared loss and a linear readout, the mean-square error of the
standard parameter-shift estimate of an empirical Fisher matrix over a data set is
an exact polynomial in the statistics of a single shot (Theorem~1). Its leading
term grows with the effective rank of the matrix, the number of independent
directions that the training samples' gradients span, a dependence that a metric of
a single state does not have. Third, the shot budget for a stated precision follows
from a short pilot run, and on the processor this protocol predicted the error of
each run and delivered the predicted saving from dividing the shots unevenly across
samples. A supporting result (Theorem~2) shows that, for the standard shift
circuits, no collective or adaptive measurement that ignores the circuit's
structure does better as the precision tightens. The effective dimension itself can
be bracketed from the data but not pinned down at a practical budget; a corrected
estimate removes most of the remaining bias and on a 96-parameter entangling model
places the effective dimension between $44.1$ and $45.1$, where the standard
estimate at 256 shots read $65.96$. The tools are standard: exact moments of
products of sample means\cite{goodman1960}, the U-statistic, and quantum
semiparametric efficiency\cite{tsang2020} with the delta method. Box~1 defines the
terms used throughout.

\begin{figure}[tbp]
\keybox{%
\textbf{Box 1. Terms used throughout.}\par\smallskip
\textit{Shot and execution.} A shot is one run of a circuit ending in a
measurement of every qubit in the computational basis. An expectation value is
estimated by averaging many shots, and its error falls as one over the square
root of their number. An execution is one shot of one circuit, and a budget
counts executions summed over all circuits.\par\smallskip
\textit{Parameter shift.} A way to measure a derivative on hardware: run the
circuit with one gate angle moved up and down by $\pi/2$ and take half the
difference of the two averages. For the gates used here the rule is exact, so the
only error is shot noise.\par\smallskip
\textit{Fisher matrix.} Here the empirical Fisher matrix: for each training
sample, take the gradient of its loss with respect to the parameters, form the
matrix of all pairwise products of its entries, and average these matrices over
the samples. Entry $(j,k)$ is large when parameters $j$ and $k$ move the loss
together on many samples.\par\smallskip
\textit{Effective dimension.} A capacity measure computed from the eigenvalues of
the normalised Fisher matrix, counting each direction by how far its eigenvalue
stands above a resolution set by the amount of data.\par\smallskip
\textit{Effective rank.} The number of directions in which the Fisher matrix is
substantially non-zero, computed from its eigenvalues as their sum squared over
the sum of their squares (the participation ratio). It is 1 when all samples'
gradients point the same way and can reach the number of samples, or of
parameters if that is smaller, when they point in unrelated
directions.\par\smallskip
\textit{Standard, split-shot and corrected estimates.} The standard estimate
averages the outer products of the estimated gradients. The split-shot estimate
divides each sample's shots into two independent halves and multiplies the gradient
of one half by that of the other, which removes the bias of the matrix. A capacity
measure is a nonlinear function of the matrix and keeps a smaller bias even then;
the corrected estimate removes most of what remains and comes with an
interval.\par\smallskip
\textit{Cumulants.} The mean, the variance and two higher statistics that
describe the asymmetry and the tails of a single shot's outcome; Theorem~1 needs
these four for every circuit.\par\smallskip
\textit{Confidence radius.} A radius computed from the same
data as the estimate, such that the true Fisher matrix lies within that distance
of the estimate with a stated probability, here 95\%, in the limit of many shots.
It is an asymptotic confidence radius, not a finite-sample guarantee, and it
covers shot noise only; its coverage at the shot counts used is checked in
simulation and on hardware. We avoid the word ``certify'', which in quantum
information usually means a finite-sample guarantee.\par\smallskip
\textit{Registered run.} A run whose prediction, pass and fail conditions and
analysis were committed to the project repository before the data were taken;
the commit history dates them, and a manifest of every plan's hash has since been
timestamped publicly (OpenTimestamps, anchored in Bitcoin), which shows that no plan
changed afterwards. Other runs are exploratory.}
\end{figure}

The closest work falls into three groups. Effective-dimension studies compare
quantum models with classical ones\cite{abbas2021,abbas2021b} and have done so on
photonic and superconducting hardware\cite{mckiernan2026}, and a recent analysis
asks what the logged quantity measures in hybrid models, naming finite shots and
device noise as open\cite{mia2026}; none of them reports the shot bias of the
measure or corrects it. Closest in aim, Hu et al.\cite{hu2023} show that finite
sampling limits the expressive capacity of physical learning machines, through
eigentasks that stay resolvable at a given shot count, and measure it on
superconducting processors, and Peng\cite{peng2026} extends a Fisher-based capacity
hierarchy for reservoir computers to finite measurement. Both ask how much capacity
survives finite sampling; we ask how far squared shot noise inflates a
gradient-based measure, how to remove the inflation, and what budget prevents it. Measurement-cost studies bound or allocate shots for
quantities built from gradients: shot-frugal
optimisers\cite{sweke2020,kubler2020,arrasmith2020}, the metric tensor of a
metric-aware variational step with an optimal split of samples between its
entries\cite{vanstraaten2021}, non-uniform shots over data points and measurement
operators\cite{moussa2023}, gradient coordinates sampled by their variance
weights\cite{harrownapp2021}, and shot distributions for quantum-geometric-tensor
quantities in imaginary-time evolution\cite{zhang2026}; we credit these for the
allocation idea. Our object is the empirical Fisher matrix of a loss over a data
set, whose cost depends on how the per-sample gradients align, a quantity that a
state-level metric does not have. Third, neighbouring estimators: unbiased
U-statistic estimates of the quantum Fisher information of a state from randomised
measurements\cite{rath2021,lusha2025}, stochastic estimates of it under
shots\cite{gacon2021}, the finite-sampling noise of a model's
output\cite{kreplin2024,kardashin2026}, the shot cost of gradients with respect to
the input\cite{li2026}, and a few-shot surrogate of the quantum Fisher information
for natural gradient\cite{naved2026}. A companion paper\cite{companion} uses the
light cones of shallow circuits to make the same matrix cheaper to measure and to
test the locality assumption from the data; here every result except the
measurement design of one run is independent of it. \VersionNote

What researchers can do differently follows directly. Anyone who reports a Fisher
spectrum or a capacity measure from hardware can state the shot budget behind it
and compute that budget from a pilot run before the experiment with the exact
formula. They can report the measure from the split-shot estimate, which needs no
noise model, rather than from the standard estimate. Anyone who compares a quantum
model with a classical one by such a measure can check, at no extra cost, whether
the comparison survives: the difference between the standard and split-shot
readings of the same data is the inflation, and a gap smaller than it is decided by
the shot count. On the model here the gap was smaller.

\section*{Results}

\subsection*{How far can shot noise move a capacity measure?}

How far can shot noise move a capacity measure computed from the Fisher matrix?
The effective dimension counts how many directions in parameter space the model
responds to, giving each eigenvalue of the normalised Fisher matrix a weight that
depends on how far it stands above a resolution set by the size of the data set.
We use its local form,
\begin{equation}
  d=\sum_{m}\frac{\log(1+k\lambda_{m})}{\log k},\qquad
  k=\frac{\gamma N}{2\pi\log N},
  \label{eq:effdim}
\end{equation}
with $\lambda_{m}$ the eigenvalues of $pF/\operatorname{tr}F$ for $p$ parameters,
$N$ data points and a constant $\gamma\in(0,1]$ that scales the data count: the effective dimension of
Ref.~\cite{abbas2021} evaluated at one parameter point instead of averaged over
parameter space. Small eigenvalues move $d$ most, and they are where shot noise
lands. In the standard estimator, which averages the outer products of the
estimated gradients, the squared noise of each estimated gradient adds a positive
term to the estimated matrix, which fills its near-zero directions and makes the
model look more expressive than it is. Figure~\ref{fig:fisher} shows the
mechanism on a two-parameter example and its size on the 32-parameter model of
the introduction ($B=64$ samples, $\gamma=1$, $N=10^{4}$), where 256 shots per
circuit raise $d$ from $14.04$ to $19.82$; it comes within $0.5$ of the exact value
only at 16,384 shots per circuit, $6.8\times10^{7}$ executions. Removing the bias of
the matrix does not remove the bias of a nonlinear function of it. The split-shot
estimator of the next section, which is unbiased for the matrix, converges with a
quarter of the shots on this model once its negative eigenvalues are set to
zero; on a blocked circuit with
64 parameters, where the exact value is $33.99$ and 256 shots per circuit read
$38.85$ with the standard estimator, it reads $31.01$ and converges no faster
(\SIeffdim). The inflation also needs room to act: with fewer samples than
parameters the estimated matrix has rank at most $B$, so shot noise cannot fill
its null directions and the inflation is capped (\SIeffdim). The size of the
inflation depends on the resolution $k$ of Eq.~\eqref{eq:effdim}, but not its
sign: on the 96-parameter model below, $N$ from $10^{3}$ to $10^{6}$ and
$\gamma\in\{0.1,1\}$ give an inflation of $1.17$ to $1.48$ at 256 shots per
circuit (\SIeffdim).

The inflation is not a simulation artefact. In a registered run on the IBM Heron
processor \device{ibm\_kingston}, the same trained model ran as eight copies on
128 qubits, each copy evaluating different inputs, with 4,096 shots for each of
its 520 circuits (4,160 evaluations of the model); subsets of the shots give the estimate at fewer shots. The
standard estimator's effective dimension read $23.42$, $19.92$, $17.10$ and
$15.29$ at 64, 256, 1,024 and 4,096 shots per circuit, against $14.49$ from the
split-shot estimator at all shots; at 256 shots that is $37\%$ above the
device's own value (registered threshold $20\%$). That reference is still
falling at 4,096 shots, so $37\%$ understates the excess; against the noiseless
value $14.04$, from which the device's matrix differs by $5\%$, it is $42\%$. In
simulation the same design read $19.82$ at 256 shots. Resampling binomial shots from the device's measured frequencies
gives $20.08$ at 256 shots, so the inflation is the size that shot noise alone
predicts. The split-shot estimator read $15.74$ at 256 shots, removing $77\%$ of
the excess (Fig.~\ref{fig:fisher}e). At a few hundred shots per circuit, then, shot noise moves a capacity measure by
tens of percent, upward when the matrix is formed in the standard way, so a
reported capacity cannot be read without the shot budget behind it.

Can the effective dimension itself be pinned down? A second registered run on
\device{ibm\_kingston} put the question to an entangling model with more inputs
than parameters: the depth-2 brickwork of CNOT gates on a line of 48 qubits
($p=96$, $B=128$ binary inputs, parity labels, noiseless value $40.78$), two
copies on two 48-qubit lines of the processor, 3,328 circuits, 4,096 shots per
circuit in the arm used here and 2,772~s of quantum time. To fit the budget, the
gradients were measured with the light-cone design of the companion
paper\cite{companion}: parameters whose light cones do not touch are shifted in the
same circuit, and each derivative is read on the qubits its gate can reach
(Methods). If the device respects the light cones, every derivative estimate is
still an unbiased two-term shift difference. The standard and split-shot estimates
are then formed from them exactly as for the coordinate-wise design, and Theorem~1
applies with the statistics of the qubits read, treating parameters shifted in the
same circuit as independent. A device need not respect the light cones, and the
run measured by how much. Derivatives read with their parameter shifted alone
differed from the jointly shifted ones by at most $0.17$ of a component's signal
(upper 95\% value). Rescaling every derivative by its measured ratio of joint to
single shift (median $0.90$) moves the effective dimension from $45.3$ to $48.8$.
The numbers below describe the matrix read on the light cones and carry this
sensitivity; the registered 16-qubit run described after the comparison reads every
derivative on all qubits and shows that the inflation does not depend on this
readout.
The standard estimator's effective dimension read $65.96$, $55.90$ and
$49.56$ at 256, 1,024 and 4,096 shots per circuit, and the split-shot estimator
$48.73$, $46.30$ and $45.32$. At 256 shots the standard value is $46\%$ above the
split-shot value at all shots, $45.30$ averaged over random halvings of the
shots (registered threshold $25\%$, simulation $52\%$). The
effective dimension of the model's own predictive distribution, from the same data,
is $45.26$, so on this model the choice between the two Fisher matrices does not
matter. To bracket $d$ we bound the eigenvalues of $F$. A bootstrap of the shot
noise gives a radius $r$ for the largest eigenvalue of $\hat F-F$ in absolute
value, and another for the trace; when both hold, every eigenvalue of $F$ lies
within $r$ of the corresponding eigenvalue of $\hat F$, which brackets $d$
(Methods). At the run's full shots the 95\% bracket is $[30.2,90.1]$; in 40
simulated runs of the design it contained the true value every time. It is wide
because $d$ is set by eigenvalues of order $1/k$ of their mean, while $r$ is $0.39$
of the mean here, and it narrows slowly: its lower end sets every eigenvalue below
$r$ to zero, so halving the shots, which raised $r$ by a factor $1.45$, widened the
half-width only from $29.9$ to $33.6$. Evaluating the bracket with $r$ scaled as
$S^{-1/2}$, shrinking it to $\pm0.5$ would take about $1.3\times10^{5}$ times the
shots.

The split-shot point estimate is far more stable than the bracket, its values from
the two halves of the run in time agreeing to $0.05$, but it is unbiased for $F$
and not for $d$: in simulation of the design it read $1.5$ above the true value.
We tried three ways to remove that remaining bias, each giving a point value and an
interval for the effective dimension of the device's matrix as read on the light
cones. The corrected estimate extrapolates the split-shot values at all, half and a
quarter of the shots, or subtracts a bootstrap estimate of the bias drawn within
each half so that the halves stay independent. It then subtracts the residual bias
that these steps leave in simulation. Its intervals covered the true value in 39
or 40 of 40 simulated runs, and on the device they place $d$ between $44.1$ and
$45.1$. That interval relies on the simulation's noise model, and the device data
suggest the model is incomplete: the two extrapolations agree to $0.02$ on the
device against $0.20$ in simulation. Two checks bear on this. With a damping of
the means added, of the kind gate noise produces and the calibration does not
know, the corrected interval still covered the true
value in 36 of 40 simulated runs, with a bias of $0.20$. A correction that needs no
noise model, a delete-a-group jackknife of the all-pairs U-statistic over the shot
chunks, gives $44.95$ \ci{44.72}{45.18} on the device and agrees with the corrected
estimate. In simulation its interval covered the true value in only 27 of 40 runs,
so we use it as a cross-check rather than as an interval (\SIeffdim). Shrinking $\hat F$ towards a multiple of the identity, the usual remedy
for a noisy covariance matrix, does not help: it treats the spread of the
per-sample gradients, which is part of what $F$ measures, as noise, and raises $d$
to $77$. A capacity measure can therefore be reported with most of its shot bias
removed, and bracketed, but on a model of this size it cannot be pinned down to a
useful tolerance at a practical budget. What does have a useful confidence radius
is $F$: $4.7\%$ in Frobenius norm from shot noise on this run, or $41\%$ once the
bound on the light-cone readout bias is included.

\begin{figure}[tbp]
\centering
\includegraphics[width=\textwidth]{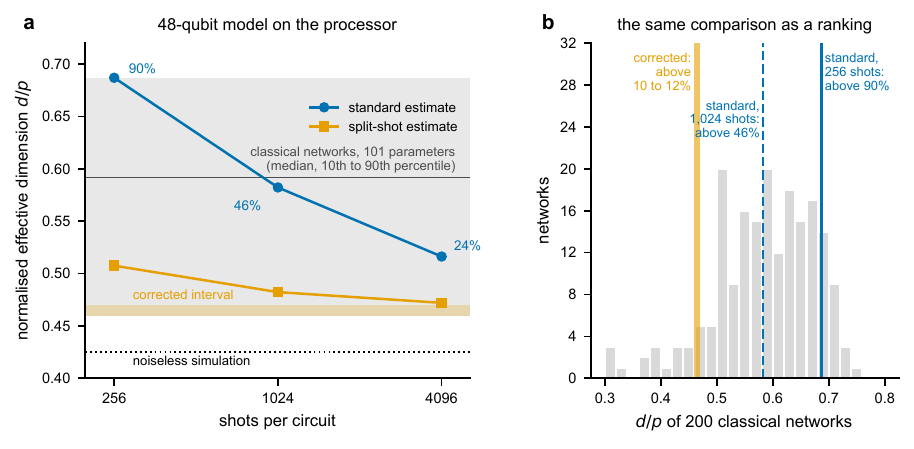}
\caption{\textbf{The shot count decides a capacity comparison.}
\textbf{a}~Normalised effective dimension $d/p$ of the 48-qubit brickwork model
($p=96$, $B=128$) on \device{ibm\_kingston}, from the standard estimate (blue) and
the split-shot estimate (orange) at 256, 1,024 and 4,096 shots per circuit, the
first two from subsets of the run's shots; the orange band is the corrected
estimate's interval, which relies on the simulation's noise model, and the dotted
line the noiseless value. Grey: $d/p$ of 200 classical networks with one hidden
layer of two units (101 parameters) on the same inputs, labels and loss, median and
10th to 90th percentile. Percentages: the share of these networks that the
standard estimate places below the quantum model. \textbf{b}~The same comparison
as a ranking: the histogram of the classical networks' $d/p$ with the quantum
model's value from the standard estimate at 256 and 1,024 shots and from the
corrected estimate. The comparison was computed after the run, on the registered
run's data.}
\label{fig:decision}
\end{figure}

\begin{figure}[tbp]
\centering
\includegraphics[width=\textwidth]{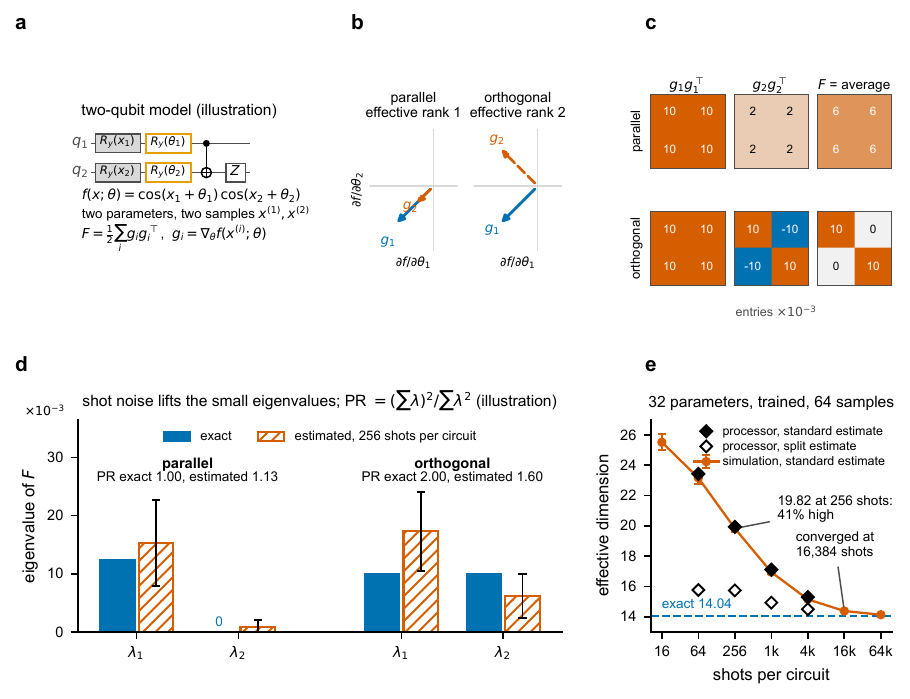}
\caption{\textbf{What the Fisher matrix is, and what shot noise does to it.}
\textbf{a}~A two-qubit model with one parameter per qubit and two training
samples. \textbf{b}~The per-sample loss gradients drawn as arrows: parallel
gradients give a matrix of effective rank 1, orthogonal gradients of the same
length give effective rank 2. \textbf{c}~The Fisher matrix as the average of the
samples' rank-one outer products. \textbf{d}~Eigenvalues of the exact matrix and
of estimates from 256 shots per circuit with the standard estimator: shot noise
spreads the eigenvalues, lifting those near zero and splitting equal ones, and
the participation ratio
$\mathrm{PR}=(\sum\lambda)^{2}/\sum\lambda^{2}$ is given under each. Panels a to d
are an illustration computed with simulated shots. \textbf{e}~Local effective
dimension of a trained 32-parameter product model ($B=64$) against shots per
circuit: exact value $14.04$, $19.82$ at 256 shots per circuit ($41\%$ high), and
within $0.5$ of the exact value only at 16,384. Diamonds: the same model on the
processor, standard estimate (filled) and split-shot estimate (open), from
subsets of 4,096 shots per circuit.}
\label{fig:fisher}
\end{figure}

The bias is large enough to decide a close comparison. We computed the comparison
after the run, on the registered run's data; it was not among the registered
checks. Classical networks with one hidden layer of two units, 101 parameters, on
the same inputs, labels and loss, have a normalised effective dimension $d/p$
between $0.46$ and $0.69$ (10th to 90th percentile over 200 random
initialisations). The quantum model's $d/p$ is $0.47$ from the split-shot estimate
at all shots, which places it above $12\%$ of these networks. It is $0.46$ to $0.47$
from the corrected estimate ($10$ to $12\%$) and $0.47$ from the jackknife ($12\%$);
rescaled by the light-cone readout ratio above, it is $0.51$ ($18\%$). From the
standard estimate at 256 shots per circuit it is $0.69$, which places the quantum
model above about $90\%$ of the networks. Within one standard deviation of that
estimate over subsets of the shots the share runs from $88$ to $91\%$, and at 1,024
shots it would be $46\%$ and at 4,096 shots $24\%$ (Fig.~\ref{fig:decision}). The $90\%$ sits
at the edge of the classical distribution, and the comparison sets one untrained
parameter point of the quantum model against many classical initialisations.
Is this reversal typical, or a property of one parameter point? We registered a
run that answers this on hardware without the light-cone readout. A 16-qubit
version of the same brickwork ($p=32$, $B=64$, parity labels) ran on
\device{ibm\_marrakesh} at six parameter points at once, on six disjoint lines of
the processor: three drawn uniformly and three trained in simulation. Every
derivative came from its own pair of shifted circuits, read on all 16 qubits, at
2,048 shots per circuit. The classical reference was a tanh network with 37
parameters, 200 random initialisations for the random points and the same networks
trained for the trained ones. At the three random points the standard estimate at
256 shots read $d=25.9$, $26.5$ and $26.8$, above every classical network, and the
split-shot estimate at all shots read $14.6$, $16.7$ and $16.2$, above $11$, $22$
and $19\%$ of them. The jackknife of the all-pairs U-statistic agreed with it
($13.4$, $16.8$, $16.2$). At the three trained points the device's matrix sat high
on either estimate: above $89$ to $95\%$ of the trained networks from the
split-shot estimate and $98$ to $100\%$ from the standard one. Noise on the device,
beyond the readout errors that the prediction includes, raised these points'
effective dimension to two to three times its noiseless value of 4 to 6, so no shot
count would have changed that comparison. All three registered
checks passed. At every point the inflation was at least half of what shot noise
alone predicts (observed $0.70$ to $1.01$ times the prediction). The mean shift of
the classical rank was $45$ percentage points against a threshold of $26$, and the
forecast described below exceeded the observed reading at every point, by $7$ to
$9\%$. We also simulated the same model at 100 random parameter points. The true
value lay below the classical median at every point, yet the standard estimate at
256 shots placed every point above it, and the split-shot estimate placed none
above it; at the 16 trained points, where the true value lay above
the median at 14, both estimates placed all 16 above it (\SIeffdim). The
parameter-averaged effective dimension of Ref.~\cite{abbas2021} moves in the same
way, from $0.49$ exact to $0.82$ from the standard estimate at 256 shots, close to
the classical networks' $0.86$. The rule a reader can apply is simple. The standard estimate exceeds
the split-shot one by the inflation, and a comparison is decided by the shot count
when the gap it reports is smaller than that. The inflation can also be forecast
before the run. The expected standard estimate is $F+D$, with $D$ the covariance of
the gradient noise, which a pilot run measures. The leading scale of the inflation
is $\operatorname{tr}D/\operatorname{tr}F$, about $pV/(2GS)$ for the
coordinate-wise design in the notation of the next section. Here
$\operatorname{tr}D/\operatorname{tr}F$ was $0.13$ at 256 shots, and $d(F+D)$ read
$74$ against the $66$ observed. The forecast errs high, as it did at every
point and shot count of the simulation across parameter points and at all six
points of the registered run, so it is a conservative screen.

\subsection*{How many shots does a stated precision need?}

How many shots does a stated precision need, and can the number be known before
the experiment? We measure precision the way the phrase ``the Fisher matrix to
within 15\%'' would be read: the root-mean-square distance between the estimated
and the true matrix, as a fraction of the size of the true matrix (its Frobenius
norm). The effective rank enters through that yardstick. To leading order, the
error that shot noise adds to the matrix depends only on the sizes of the
gradients and on the readout noise, while the size of the matrix also depends on
how the gradients are oriented. Take two qubits, one parameter on each, and two
training samples with gradients of equal length. If the two gradients are
parallel, the matrix has a single direction, effective rank 1, and its Frobenius
norm equals its trace. If they are orthogonal, the same trace is shared by two
equal eigenvalues, the effective rank is 2, the norm falls by a factor of
$\sqrt{2}$, and the same absolute error becomes a relative error $\sqrt{2}$ times
larger, so twice the shots are needed (Fig.~\ref{fig:fisher}b). Theorem~1 makes
this exact for any circuit measured by two-term parameter shift, under squared loss
with a linear readout; for a log-loss, or for a classifier's Fisher matrix, which
divides by the predicted probability, the error is not a polynomial and the
split-shot estimator is no longer exactly unbiased.

Write the model output as $f_{\theta}(x)=\tfrac{1}{n}\sum_{q}\langle
Z_{q}\rangle$, the mean Pauli-$Z$ readout of $n$ qubits, the squared loss
$L=(f-y)^{2}$, and the empirical Fisher matrix
$F=\tfrac{1}{B}\sum_{i}g_{i}g_{i}^{\mathsf{T}}$ over a minibatch of $B$ samples,
with $g_{i}=\nabla_{\theta}L_{i}=2(f_{i}-y_{i})\nabla_{\theta}f_{i}$ and $p$
parameters. Four measurable quantities carry the cost.
$V=n^{-2}\sum_{q,q'}\operatorname{Cov}(Z_{q},Z_{q'})$ is the variance of a
single-shot estimate of $f$; it sums the full covariance of the measured qubits,
off-diagonal entries included, so entanglement enters through it.
$G=\lVert\nabla_{\theta}f\rVert^{2}$ is the squared gradient norm, the signal.
The alignment $\kappa=\lVert F\rVert_{F}\big/\,\overline{\lVert
g_{i}\rVert^{2}}$, where the bar is the mean over samples, lies between
$\min(B,p)^{-1/2}$ for orthogonal gradients and 1 for parallel ones. And $h\ge1$ measures how unevenly the residuals and signals
are spread across samples, with $h=1$ when they are even (Methods). Each sample
receives its own shots: $S$ executions at $\theta$ and at each of the $2p$
shifted configurations, so $\BQ=B(1+2p)S$ executions in total. With $S$
executions per configuration the two-term parameter-shift estimate of
$\partial_{j}f$ has variance $w_{j}/S$, where
$w_{j}=[V(\theta+\tfrac{\pi}{2}e_{j})+V(\theta-\tfrac{\pi}{2}e_{j})]/4$, which is $V/2$
when the shifted configurations have the variance of the unshifted one (Lemma~A,
Methods).

\runin{Theorem 1 (the exact risk).}
Under coordinate-wise parameter shift with independent shots for every sample and
configuration, the per-sample gradient estimate is unbiased and the mean-square
error $\mathbb{E}\lVert\hat F-F\rVert_{F}^{2}$ is an exact polynomial in $1/S$
determined by the first four single-shot cumulants of $f$ at the $1+2p$
configurations of every sample (Methods, Eq.~\eqref{eq:exact}). To leading order,
\begin{equation}
  \frac{\mathbb{E}\lVert\hat F-F\rVert_{F}^{2}}{\lVert F\rVert_{F}^{2}}
  \simeq\frac{pV}{\kappa^{2}GS}\left(\frac{h}{B}+\frac{V}{4GS}\right).
  \label{eq:cost}
\end{equation}
Reaching relative mean-square error $\eps^{2}$ therefore needs
\begin{equation}
  S=\Theta\!\left(\frac{pVh}{\eps^{2}\kappa^{2}GB}
      +\frac{\sqrt{p}\,V}{\eps\kappa G}\right),
  \qquad \BQ=B(1+2p)\,S .
  \label{eq:shots}
\end{equation}
The second term is a bias of order $1/S$ from the squared shot noise, the term
that inflates the effective dimension above. It is removable at no cost:
splitting each sample's shots into two independent halves and estimating
$\tfrac{1}{B}\sum_{i}\tfrac12(\hat g_{i}^{(1)}\hat
g_{i}^{(2)\mathsf{T}}+\hat g_{i}^{(2)}\hat g_{i}^{(1)\mathsf{T}})$ gives an
unbiased estimator with the same leading fluctuation (Methods). We call it the
split-shot estimator, and the one that averages $\hat g_{i}\hat g_{i}^{\mathsf T}$
the standard estimator.

\runin{Lemma (alignment is effective rank).}
Since $\operatorname{tr}F=B^{-1}\sum_{i}\lVert g_{i}\rVert^{2}$,
\begin{equation}
  \kappa^{2}=\frac{\lVert F\rVert_{F}^{2}}{(\operatorname{tr}F)^{2}}=\frac{1}{\mathrm{PR}(F)},
  \qquad \mathrm{PR}(F)=\frac{(\sum_{k}\lambda_{k})^{2}}{\sum_{k}\lambda_{k}^{2}},
  \label{eq:kappapr}
\end{equation}
the inverse of the participation ratio of the Fisher spectrum, and while the first term
of Eq.~\eqref{eq:shots} dominates, which holds for $B\lesssim2h\sqrt{p}/(\eps\kappa)$,
\begin{equation}
  \BQ\simeq 2h\,p^{2}\,\mathrm{PR}(F)\,\frac{V}{\eps^{2}G},
  \label{eq:prlaw}
\end{equation}
the factor 2 from the two shifted configurations per parameter. The explicit
minibatch factor cancels, and $B$ enters only through the effective
rank: $1\le\mathrm{PR}(F)\le\min(B,p)$, so samples whose gradients point in new
directions raise the cost and samples whose gradients repeat old ones do not. In
the two-qubit example, $\mathrm{PR}(F)$ is 1 for parallel and 2 for orthogonal
gradients, and Eq.~\eqref{eq:prlaw} doubles.

\runin{What the circuit fixes, and what it does not.}
The prefactor $V/G$ of Eq.~\eqref{eq:prlaw} is set by how far each gate's influence
spreads, its light cone. If every Heisenberg-evolved $Z_{q}$ is supported on at
most $w$ qubits, then $\operatorname{Cov}(Z_{q},Z_{q'})=0$ for disjoint supports
and $\partial_{j}\langle Z_{q}\rangle=0$ unless qubit $q$ lies in the light cone of
gate $j$, so $nV$ and $nG$ are bounded independently of the register size and
$V/G$ stays fixed as qubits are added (eleven families in \SIfixed). The effective
rank is not fixed by the circuit. On circuits of disjoint blocks it grows with the
number of blocks until it nears the minibatch size, so a cost exponent fitted to one
family is a property of the data and the minibatch as well as of the circuit;
\SIfixed{} gives a phenomenological fit of this growth and the registered sweeps,
in which the exponent of the budget in $p$ rose from $2.12$ to $2.74$ as $B$ grew.

\runin{The formula predicts measured cost without fitted constants.}
The exact mean-square error agrees with 4,000-repetition Monte Carlo in all 12
cells tested (no deviation beyond $1.29$ standard errors), and the split-shot
estimator with its own expression in all four (largest $|z|=1.31$). The
leading-order expression~\eqref{eq:cost} is within $0.6\%$ of the exact value at
$S=1024$ and up to $26\%$ low at $S\le64$. On a product family to 256 qubits
Theorem~1 predicts an executions exponent in $p$ of $1.965$, against
$1.966$ \ci{1.934}{1.997} measured, and on blocked circuits to 128 qubits a
matched-precision shot search gives exponents of $2.27$ and $3.04$ against $2.26$
and $3.02$ predicted (\SIfixed). The yardstick decides whether the effective rank
enters at all. Measured relative to $\operatorname{tr}F$ it drops out,
$\BQ^{\mathrm{tr}}\simeq2h\,p^{2}V/(\eps^{2}G)$; quantities of
$F/\operatorname{tr}F$ such as the effective dimension need only this
trace-relative error, while a preconditioner or a statement about individual
directions needs the Frobenius-relative one (Methods gives budgets in both). In
either case the shots for a stated precision follow from Theorem~1 before the run.

\subsection*{Can a better measurement or a better division of shots do better?}

Once the circuits are fixed, can a better measurement or a better division of the
shots lower the cost? The intuition for the first part is that the Fisher matrix
depends on each circuit's output state only through the mean of one observable,
the average $Z$ readout, and that observable is diagonal in the computational
basis. Reading every qubit in that basis is then the most informative thing a
measurement of one copy can do: any single-copy measurement that estimates the
mean without bias at every state has at least the variance of $f$ itself, which
the computational-basis readout attains (Methods). Theorem~2 shows that measuring
many copies jointly, or choosing later measurements from earlier outcomes, adds
nothing at high precision either.

\runin{Theorem 2 (efficient within the unstructured class).}
Let every sample $i$ contribute the $1+2p$ configuration states of the
parameter-shift design, with means $\mu_{i0}=f(\theta;x_{i})$ and
$\mu_{ij\pm}=f(\theta\pm\tfrac{\pi}{2}e_{j};x_{i})$, and let each state be
otherwise unrestricted: the local model around each configuration state is the
full state space, with no structural knowledge (no known product structure, no
known light cone). For any strategy that spends $N_{c}=\nu_{c}N$ copies (preparations of a circuit's
output state) on configuration $c$, whether adaptive or with collective measurements across
copies, and for any estimator $\hat F$, the local asymptotic minimax risk at fixed
$n$ obeys
\begin{equation}
\begin{gathered}
  \lim_{s\to\infty}\liminf_{N\to\infty}\,N\!\sup_{U_{s/\sqrt N}}\mathbb{E}\lVert\hat F-F\rVert_{F}^{2}
  \\ \;\ge\;\sum_{c}\frac{A_{c}V_{c}}{\nu_{c}},\\
  A_{c}=\Bigl\lVert\frac{\partial F}{\partial\mu_{c}}\Bigr\rVert_{F}^{2},\qquad
  V_{c}=\operatorname{Var}_{\rho_{c}}(f),
\end{gathered}
  \label{eq:lb}
\end{equation}
where $U_{r}$ is the set of configuration states within $r$ of the true ones
along the curves of Methods, the supremum is over the states in $U_{r}$, and
$A_{i0}=4\lVert g_{i}\rVert^{4}/(B^{2}r_{i}^{2})$,
$A_{ij\pm}=r_{i}^{2}(2\lVert g_{i}\rVert^{2}+2g_{ij}^{2})/B^{2}$, $r_{i}=f_{i}-y_{i}$.
Counts chosen from earlier readouts, as in the pilot-then-allocate protocol
below, are covered with $\mathbb{E}N_{c}$ in place of $N_{c}$ (Methods). In
words, the left side is the worst mean-square error, scaled by the number of
executions, over states in a shrinking neighbourhood of the true ones; no
measurement strategy and no estimator can push it below the right side.
The fewest executions that reach relative mean-square error $\eps^{2}$ are
\begin{equation}
  N^{\star}(\eps)=\frac{\bigl(\sum_{c}\sqrt{A_{c}V_{c}}\bigr)^{2}}{\eps^{2}\lVert F\rVert_{F}^{2}}\,(1+o(1)),
  \label{eq:nstar}
\end{equation}
attained by computational-basis readout with $N_{c}\propto\sqrt{A_{c}V_{c}}$. The
leading fluctuation of Theorem~1 equals the right-hand side of Eq.~\eqref{eq:lb}
term by term, so the split-shot estimator is asymptotically efficient within the
class: given the shift configurations, no collective or adaptive measurement and
no estimator that ignores structure does better as $\eps\to0$. At fixed $n$ the
standard estimator is efficient as well, since its squared bias is of order
$1/S^{2}$ and drops out of the scaled limit. The bound therefore ranks designs and
allocations rather than the two estimators. Their difference for a capacity
measure is the bias term of Eq.~\eqref{eq:cost} at the shot counts used. It also concerns the
Frobenius error of $F$, not the error of $d$. In the homogeneous
reading $N^{\star}\simeq 2p^{2}V/(\eps^{2}\kappa^{2}G)$, and uniform shots cost
about $h$ times more, exactly so up to $O(1/p)$ when the per-sample signals
$G_{i}$ and the variances $V_{c}$ are homogeneous. The proof (Methods) uses a
curve through each configuration state that stays a state, pure when the state
is pure, together with the Braunstein--Caves inequality and the multivariate van
Trees inequality. Exact split-estimator risk over the bound is $1.20$, $1.05$,
$1.012$, $1.003$ and $1.0002$ at $S=16$, 64, 256, 1024 and 16,384 shots per
configuration on a product model (\SIfixed).

The class is stated deliberately. Estimators that do not query the shift
configurations one parameter at a time, or that use structure, lie outside it:
random-direction estimation of the kind used by simultaneous-perturbation
stochastic approximation (SPSA), estimators that read the whole gradient from a
few settings when the generators commute\cite{bowles2025,chinzei2025}, and
estimators that use a known light cone\cite{companion}. Methods lists the others.
Within the standard shift circuits, further savings therefore need something the
class leaves out: structure of the circuit, other configuration sets, control
variates or other observables.

\runin{Dividing the shots.}
The division that matters is across samples. Allocating per coordinate,
$S_{j}\propto\sqrt{a_{j}}$ (Methods), saves at most $1.4\%$ over uniform on every
family studied here. Across samples, where shots have been allocated non-uniformly
before\cite{moussa2023}, minimising the exact risk saves $1.6$ to $2.6\times$ in
median per family ($1.1$ to $7.5\times$ across cells); $h$ is the price of uniform
shots. The bound is asymptotic, and reaching for it naively fails: Neyman
allocation built on the leading term alone starves the unshifted configuration of
well-fitted samples, where $r_{i}$ is small, the second-order term
$v_{i}w_{ij}/(S_{i0}S_{ij})$ that Eq.~\eqref{eq:lb} drops then dominates, and on a
trained product model the error overshoots its target $\eps^{2}=0.01$ 87-fold.
Minimising the exact risk of Theorem~1 instead lands within $3$ to $8\%$ of
$N^{\star}$ (median, 48 product and chain cells at $\eps=0.1$) and meets the target
in Monte Carlo (relative risk $0.0096$ to $0.0104$). On the blocked circuit to 64
qubits Monte Carlo matches the exact prediction within $2\%$ in all 12 arms, and the
exact-risk allocation needs $1.61$ to $4.44$ times fewer executions than uniform
shots, $1.2$ to $2.8\%$ above $N^{\star}$ (\SIfixed).

\runin{A confidence radius from the data.}
The budget needs $V$, $G$ and $\kappa$, that is, the answer, to price the answer.
A plug-in confidence radius removes the circularity: inserting the measured means and
per-configuration variances into the linearised error $\sum_{c}J_{c}e_{c}$,
$e_{c}\sim\mathcal N(0,\hat V_{c}/N_{c})$, and taking the $1-\delta$ quantile of
its Frobenius norm gives a radius $\hat\rho$ with
$\Pr(\lVert\hat F-F\rVert_{F}\le\hat\rho)\to1-\delta$ (by the delta method and
Slutsky's theorem; a doubling rule that stops when
$\hat\rho\le\eps\lVert\hat F\rVert_{F}$ is valid by Anscombe's theorem). At nominal
$95\%$ its coverage is $0.935$ to $0.965$ on product models and $0.939$ on the
blocked circuit to 64 qubits (1,800 trials), one percentage point below nominal; higher-order
terms add only $1\%$ to the risk there, which points to the plug-in Jacobian. Given
the shift circuits and no structural knowledge, then, no measurement does better
at high precision; what remains is to divide the shots by the exact risk and to
report the radius.

\subsection*{Does the protocol hold up on a quantum processor?}

Does the protocol hold up on a quantum processor, where gate noise, drift and
crosstalk are present? The protocol has four steps: list the
circuits; measure each circuit's single-shot statistics in a pilot job; turn them
into a shot allocation and a predicted error with the exact risk; and estimate the
matrix in independent replicate blocks, each with its own radius. When the budget
cannot be met, a fifth step reports a capacity measure from the split-shot or
corrected estimate with its bracket, as in the runs above. In the registered runs
every prediction was fixed in a plan committed to the project repository before the
jobs were submitted. Table~\ref{tab:hardware} lists every hardware run of this
paper.

\begin{figure}[tbp]
\centering
\includegraphics[width=0.78\textwidth]{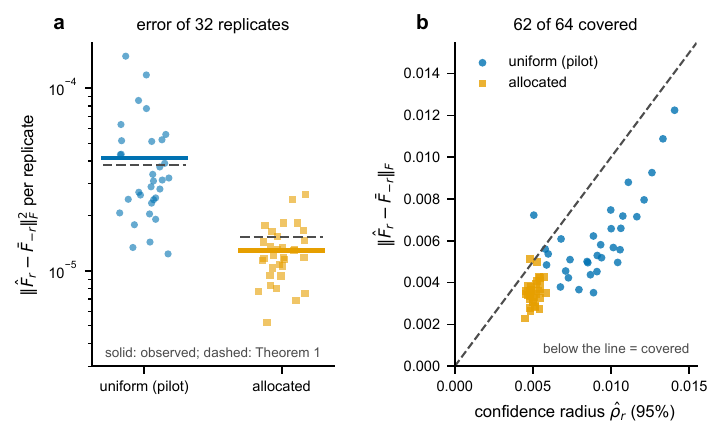}
\caption{\textbf{The protocol on the processor.}
\textbf{a}~Squared Frobenius distance of each of 32 replicate estimates
from the mean of the other 31, for the uniform pilot job (blue) and the job
allocated from it (orange), registered run on a six-qubit line of
\device{ibm\_marrakesh}; solid bars are observed means and dashed bars the
prediction of Theorem~1 at the hardware moments.
\textbf{b}~The same distance against each replicate's confidence radius at
nominal 95\%; points below the dashed line are covered (62 of 64).}
\label{fig:hardware}
\end{figure}

On a trained six-qubit chain model ($p=12$, $B=8$, 200 circuits), two jobs of
409,600 shots each tested steps 2 to 4 with nothing fitted to the device: the first
spent shots uniformly and served as the pilot, the second used the exact-risk
allocation computed from it, at the same total. The exact risk predicted the pilot
job's replicate error to a ratio of $1.09$ \ci{0.84}{1.38}; the allocation, fixed
before the second job ran, reduced the error $3.20$-fold \ci{2.42}{4.17} against
$2.72$ predicted; and the radius covered 62 of 64 replicates at nominal 95\%
(Fig.~\ref{fig:hardware}a,b). In the allocated job the exact risk over-predicts the
error by about 15\% ($0.85$ \ci{0.75}{0.96}). The interval excludes 1, so the
prediction is biased on that job; it uses the job's own moments, so drift between
the jobs does not explain it, and we have not identified the cause.
At this size the device costs little: the efficient budget from hardware moments
is $1.11$ times the noiseless one, while the hardware $\lVert F\rVert_{F}$ lies
$10$ to $18\%$ below the noiseless value. What is estimated is the device's Fisher
matrix, with its gate and readout noise, which is the matrix that an optimiser or a
capacity measure computed on this device sees.

\begin{table}[tbp]
\centering\footnotesize
\caption{\textbf{Every hardware run of this paper.} The line runs are on
\device{ibm\_marrakesh} and the capacity-measure runs on \device{ibm\_kingston};
all are registered unless marked. Error, obs./pred.: replicate error over the
exact-risk prediction (pilot job; allocated job). Coverage: replicate blocks inside
their 95\% radius. Allocation: error reduction of the allocated job at equal shots,
with the prediction fixed before the job in parentheses. Inflation: standard
estimate of the effective dimension at 256 shots per circuit over the split-shot
estimate at all shots, with the registered threshold in parentheses. A dash marks
a quantity not measured. $^{a}$Weighting the readout by the weights of the error formula
saved $0.68$ \ci{0.48}{0.98}, registered to exceed 1. $^{b}$Eight copies of a
16-qubit model without entangling gates; its three registered checks passed.
$^{c}$Two copies of a 48-qubit CNOT brickwork, gradients measured with the
light-cone design of the companion paper\cite{companion}; the three registered
checks of this paper passed, and the three that concern the light-cone design are
reported there.}
\label{tab:hardware}
\setlength{\tabcolsep}{4pt}
\begin{tabular}{@{}>{\raggedright\arraybackslash}p{4.2cm}>{\raggedright\arraybackslash}p{1.2cm}>{\raggedright\arraybackslash}p{2.0cm}>{\raggedright\arraybackslash}p{1.4cm}>{\raggedright\arraybackslash}p{2.2cm}>{\raggedright\arraybackslash}p{2.6cm}@{}}
\toprule
Run & Qubits & Error, obs./pred. & Coverage & Allocation & Inflation of $d$\\
\midrule
Line, pilot then allocation & 6 & $1.09$; $0.85$ & 62/64 & $3.20$ ($2.72$) & --\\
Line, same, exploratory & 6 & $0.86$; $0.97$ & 63/64 & $2.39$ ($2.77$) & --\\
Line, readout weights$^{a}$ & 6 & $1.07$; $1.06$ & 60/64 & $3.18$ ($3.75$) & --\\
Capacity measure, product model$^{b}$ & 128 & -- & -- & -- & $37\%$ ($20\%$)\\
Capacity measure, entangling$^{c}$ & 96 & $0.99$ & 16/16 & -- & $46\%$ ($25\%$)\\
\bottomrule
\end{tabular}
\end{table}

On the 96-qubit capacity run the same estimate of $F$ with its radius held at
scale, for the matrix read on the light cones: the radius covered 16 of 16
replicate blocks and the exact risk predicted the replicate error to a ratio of
$0.99$. Across the runs, then, the observed error was $0.85$ to $1.09$ times the
predicted one, with intervals reaching from $0.70$ to $1.38$ and one job's interval
excluding 1; the allocation computed from a pilot delivered the predicted saving,
and the radius covered at its nominal level. What the radius cannot cover is the
bias of a nonlinear function of $F$, which the split-shot and corrected estimates
address.

\section*{Discussion}

A capacity measure computed from a hardware Fisher matrix is a function of the shot
count as much as of the model. Below the budget of Theorem~1 its shot bias is of
the same order as the quantity reported, $41\%$ in simulation and $37\%$ and $46\%$
on the processor, and on the 48-qubit model it was large enough to move the quantum
model, in a close comparison, from above most classical networks of the same size
to above an eighth of them.
Comparisons of quantum and classical models by such measures, made on hardware or
with simulated shots, should therefore be read together with the shot count behind
them. Where that count is not stated, the comparison cannot be checked.

Removing the bias is much easier than pinning down the value. The split-shot
estimate, which costs nothing beyond keeping each sample's shots in two halves,
removed most of the inflation on both hardware models, and the corrected estimate
removed most of what remained. Pinning down the value is a different matter: on the
96-parameter run the data alone bracketed the effective dimension only between 30
and 90, and a tolerance of $\pm0.5$ needs the eigenvalues of $F$ to a small
fraction of $1/k$ of their mean, about $10^{5}$ times that run's shots. We
therefore recommend three things. Report a Fisher matrix or a capacity measure from
hardware with its shot budget and its error radius. Report a capacity measure from
the split-shot estimate, with the corrected value and the bracket beside it, not as
an exact value. And before comparing models by such a measure, check that the
reported gap exceeds the difference between the standard and split-shot readings.

The budget itself can be known before the experiment. Theorem~1 gives the
mean-square error exactly from four single-shot statistics per circuit, which a
pilot job measures, and on the processor the observed error of every run was
$0.85$ to $1.09$ times its prediction, with nothing fitted to the device. To leading order the budget
depends on the parameter count, the ratio of readout variance to signal, and the
effective rank of the matrix. The last of these is the part that is easy to
overlook: it depends on the data and the minibatch, not only on the circuit, so a
cost exponent fitted to one circuit family moved from $2.12$ to $2.74$ as the
minibatch grew. We recommend reporting cost through the scalings of $V$, $G$ and
$\mathrm{PR}(F)$, from which any fitted exponent follows. The \device{shotcost}
tool in the reproduction archive computes the budget and the allocation from a
pilot run.

Without structural knowledge of the circuit, the standard shift design with
exact-risk allocation is already the best one can do: given the shift
configurations, no collective or adaptive measurement and no estimator that ignores
structure does better asymptotically (Theorem~2). Further savings must come from
outside that class, above all from structure. The companion
paper\cite{companion} uses the light cones of shallow circuits for that, at the
price of an assumption about the device that it tests from the data. A second
lesson is common to both papers. A fast or convenient estimate of a nonlinear
quantity can be biased in ways that repetition does not reveal: here the bias
comes from squared shot noise, and only an estimator built to remove it, or a
budget computed to make it negligible, controls it.

What the runs estimate is the device's Fisher matrix, with its gate and readout
noise, which is the matrix that an optimiser or a capacity measure computed on this
device sees; on the six-qubit line its norm lay 10 to 18\% below the noiseless
value and on the 96-qubit run 17\%. The circuits run here are small enough or
shallow enough to simulate classically, which is how every prediction was computed
before the data; we claim no classical hardness. The protocol and the corrections do
not depend on simulability.

The results have limits. Most simulated cells use independently drawn parameters
and binary labels. On models trained on two real data sets the leading-order law
held as stated, but finite-shot terms moved the exact budgets, and all three
registered checks failed there: the budget per unit effective rank rose
$1.56$-fold from $B=16$ to 256 (leading order $1.02$), and training lowered the
budget threefold through $V/G$ rather than through the effective rank, which stayed
between $1.2$ and $7.3$ (\SItrained). The measured scaling stops at 256 qubits, on
one-dimensional families at depth 2. Theorem~2 holds at fixed $n$ in the local
asymptotic minimax sense and concerns the standard shift circuits. The confidence
radius and the bracket for the effective dimension are asymptotic; the radius ran
one percentage point below nominal on the blocked circuit, the bracket's coverage
was checked in simulation only, and a finite-sample radius, for example from a
matrix Bernstein inequality, is not derived here. The corrected estimate's interval
relies on the simulation's noise model, which the jackknife that needs none supports
but does not replace. The 96-parameter run's numbers describe the matrix read on
the light cones. The capacity comparison on that run, the first claim of the
abstract, was computed after the run and was not registered. The hardware runs are on two processors of one family. For a
readout that needs two measurement bases, Theorem~1's exact risk holds with each
configuration's variance replaced by the two bases' variances, each divided by its
share of the shots; it matched Monte Carlo to within $1\%$ on such a readout
(\SIfixed). Registration is to the project repository, dated by its commit history;
the public timestamp of the plans came after these runs.

\section*{Methods}

\subsection*{Notation and standard tools}

\SInotation{} lists every symbol. Several ingredients are standard, and we use
them as such: the exact moment algebra of products of sample
means\cite{goodman1960}, carried to fourth order; the
U-statistic\cite{hoeffding1948} behind the split-shot estimator; Neyman
allocation\cite{neyman1934}; and the effective rank as the dimension that governs
covariance estimation\cite{koltchinskii2017}. Shift rules for higher derivatives and the
quantum Fisher information are given by Mari et al.\cite{mari2021}.

\subsection*{Circuits and observables}

Three input conventions are used, and each result states which. The scale
families (brickwork, chain, ring and blocked circuits at $n\ge16$, and the
measured blocked arms) encode $x_{q}\in\{0,\pi\}$ as $R_{X}(x_{q})$, a
computational-basis bit flip, then apply \texttt{depth} layers of
$R_{X}(\theta)$ on every qubit followed by an entangler; labels are $\pm1$
parity. The small-$n$ entangling circuits used for the cost-law constants and
the minibatch sweep encode each bit as $R_{X}(\pi x_{q})$ with
$x_{q}\in\{0,\pi\}$, a rotation by $0$ or $\pi^{2}$, which flips the qubit with
probability $\sin^{2}(\pi^{2}/2)\approx0.95$. This is a valid encoding, but a near
bit flip rather than the exact one intended, an artefact of an early convention.
We kept it rather than rerun these cells because what they test does not depend on
the encoding: the checks of Theorem~1 hold for any fixed circuit, and the $V$, $G$
and $\kappa$ quoted for these circuits are those of this encoding. The effective-rank sweeps and the hardware runs encode $R_{X}(x_{q})$. The product
family is simulated exactly from single-qubit Bloch vectors, with
$R_{Y}(\theta)$ followed by $R_{Z}(x_{q})$ in every layer, starting from
$|0\rangle^{\otimes n}$. Every variational angle enters exactly one rotation with
a $\pm\tfrac12$ generator spectrum, so the two-term shift rule is exact.

Four entanglers are used. \texttt{chain} applies $\mathrm{CNOT}(q,q{+}1)$ for
$q=0\ldots n{-}2$ sequentially. \texttt{ring} adds the wrap
$\mathrm{CNOT}(n{-}1,0)$. \texttt{brickwork} applies even pairs then odd pairs,
which is parallel and has a light cone of $2d+1$ at depth $d$. \texttt{blocked}
applies the chain inside disjoint blocks of length $L$ and never across a block
boundary, so the light cone is exactly $L$ at any depth. The measured
observable is the uniform mean $f=\tfrac{1}{n}\sum_{q}\langle Z_{q}\rangle$
throughout, and gradients use the two-term parameter-shift rule with $\pm\pi/2$
shifts.

\subsection*{Proof of Theorem 1}

\runin{Lemma A.} Let a configuration $c$ have single-shot outcome
distribution of $f$ with cumulants $k_{1}^{c},\dots,k_{4}^{c}$. The mean of $S$
independent shots has cumulants $k_{r}^{c}/S^{r-1}$. The shift estimate
$\hat b_{j}=[\hat f(\theta+\tfrac{\pi}{2}e_{j})-\hat f(\theta-\tfrac{\pi}{2}e_{j})]/2$
is unbiased for $\partial_{j}f$, has variance
$w_{j}/S=[V(\theta+\tfrac{\pi}{2}e_{j})+V(\theta-\tfrac{\pi}{2}e_{j})]/(4S)$, and is
independent of $\hat b_{k}$ for $k\neq j$ because each shifted circuit is executed
separately. Its cumulants of order $r$ are
$[k_{r}^{+j}+(-1)^{r}k_{r}^{-j}]/(2^{r}S^{r-1})$.

\runin{Exact mean-square error.} For sample $i$ write
$\hat a_{i}=\hat f(\theta;x_{i})-y_{i}$ and $\hat g_{ij}=2\hat a_{i}\hat b_{ij}$,
with $\hat a_{i}$ independent of the $\hat b_{ij}$ and all samples independent.
Then $\hat F_{jk}=\tfrac{4}{B}\sum_{i}\hat a_{i}^{2}\hat b_{ij}\hat b_{ik}$ and,
exactly,
\begin{equation}
\begin{aligned}
\mathbb{E}\hat F_{jk}&=\frac{4}{B}\sum_{i}\mathbb{E}[\hat a_{i}^{2}]\,
  \mathbb{E}[\hat b_{ij}\hat b_{ik}],\\
\operatorname{Var}\hat F_{jk}&=\frac{16}{B^{2}}\sum_{i}\Bigl(
  \mathbb{E}[\hat a_{i}^{4}]\,\mathbb{E}[\hat b_{ij}^{2}\hat b_{ik}^{2}]\\
  &\qquad-\mathbb{E}[\hat a_{i}^{2}]^{2}\,\mathbb{E}[\hat b_{ij}\hat b_{ik}]^{2}\Bigr),\\
\mathbb{E}\lVert\hat F-F\rVert_{F}^{2}&=\sum_{j,k}\Bigl(\operatorname{Var}\hat F_{jk}\\
  &\qquad+(\mathbb{E}\hat F_{jk}-F_{jk})^{2}\Bigr),
\end{aligned}
\label{eq:exact}
\end{equation}
where $\mathbb{E}[\hat b_{ij}\hat b_{ik}]$ factorises for $j\neq k$ and every
moment follows from Lemma~A. The result is a polynomial in $1/S$, of degree at
most six for the standard estimator, whose variance contains the product
$k_{4}^{a}k_{4}^{b}/S^{6}$ of two fourth cumulants, and at most four for the
split-shot estimator. The precision criterion is
$\mathbb{E}\lVert\hat F-F\rVert_{F}^{2}\le\eps^{2}\lVert F\rVert_{F}^{2}$, which by
Jensen's inequality implies $\mathbb{E}\lVert\hat F-F\rVert_{F}\le\eps\lVert
F\rVert_{F}$; the required $S$ is found by bisection. With per-sample or
per-configuration shot counts the same expressions hold with $S$ replaced by the
count of each configuration, which is what the exact-risk allocation minimises.

\runin{Leading order.} Write $\delta_{i}=\hat g_{i}-g_{i}$ and
$C_{i}=\operatorname{Cov}\hat g_{i}=4[(v_{i}/S)\,d_{i}d_{i}^{\mathsf{T}}
+\operatorname{diag}((r_{i}^{2}+v_{i}/S)\,w_{ij}/S)]$, with $r_{i}=f_{i}-y_{i}$,
$d_{i}=\nabla_{\theta}f_{i}$ and $v_{i}$ the single-shot variance at $\theta$.
Expanding
$\hat F-F=\tfrac1B\sum_{i}(\delta_{i}g_{i}^{\mathsf{T}}+g_{i}\delta_{i}^{\mathsf{T}})
+\tfrac1B\sum_{i}\delta_{i}\delta_{i}^{\mathsf{T}}$, the second sum has mean
$D=\tfrac1B\sum_{i}C_{i}$, a bias of order $1/S$. The fluctuation is a sum of $B$
independent zero-mean matrices, so
\begin{multline*}
\mathbb{E}\lVert\hat F-F\rVert_{F}^{2}=\lVert D\rVert_{F}^{2}\\
+\frac{1}{B^{2}}\sum_{i}\bigl(2\lVert g_{i}\rVert^{2}\operatorname{tr}C_{i}
+2g_{i}^{\mathsf{T}}C_{i}g_{i}\bigr)\\+O\!\left(\frac{1}{BS^{2}}\right).
\end{multline*}
With $w_{ij}\approx V/2$ and $\operatorname{tr}C_{i}\approx2pr_{i}^{2}V/S$, the
fluctuation term relative to $\lVert F\rVert_{F}^{2}=16\kappa^{2}
(\overline{r^{2}G})^{2}$ is $pVh/(\kappa^{2}GBS)$ with
$h=\bar G\,\overline{r^{4}G}/(\overline{r^{2}G})^{2}\ge1$ by the Cauchy--Schwarz
inequality, and the bias term is $pV^{2}/(4\kappa^{2}G^{2}S^{2})$ to the same
order, which gives Eq.~\eqref{eq:cost}. Bounding the fluctuation by the triangle
inequality,
$\lVert\hat F-F\rVert_{F}\le\tfrac2B\sum_{i}\lVert g_{i}\rVert\lVert\delta_{i}\rVert$,
would discard the independence of the $\delta_{i}$ and with it the factor $1/B$,
giving a budget linear in $B$.

\runin{Allocation across coordinates.} Giving coordinate $j$ its own $S_{j}$,
the leading fluctuation term is $\sum_{j}a_{j}/S_{j}$ with
$a_{j}=\tfrac{8}{B^{2}}\sum_{i}r_{i}^{2}w_{ij}(\lVert g_{i}\rVert^{2}+g_{ij}^{2})$.
Minimising $\sum_{j}S_{j}$ at fixed $\sum_{j}a_{j}/S_{j}$ gives
$S_{j}\propto\sqrt{a_{j}}$ by the Cauchy--Schwarz inequality, with saving
$p\sum_{j}a_{j}/(\sum_{j}\sqrt{a_{j}})^{2}$ over uniform allocation.

\runin{Split-shot estimator.} With two independent half-budget estimates per
sample, $\hat g_{i}^{(1)}$ and $\hat g_{i}^{(2)}$, the expectation of
$\hat g_{i}^{(1)}\hat g_{i}^{(2)\mathsf{T}}$ is $g_{i}g_{i}^{\mathsf{T}}$, so
the symmetrised product is unbiased. Its fluctuation is carried by
$\bar\delta_{i}=(\delta_{i}^{(1)}+\delta_{i}^{(2)})/2$, whose covariance equals
$C_{i}$ at leading order, so the leading fluctuation is unchanged and the bias
term is removed. Its exact mean-square error follows from Lemma~A in the same way.

\runin{Checks.} Equation~\eqref{eq:exact} was compared with 4,000-repetition
Monte Carlo of the estimator on the product family ($n=6$) and the
chain ($n=5$) at $B\in\{4,32\}$ and $S\in\{16,64,1024\}$, and the split-shot
expression on the chain at four cells; every cell agrees within $1.31$ standard
errors. A second Monte Carlo on synthetic product-Bernoulli outcome laws ($n=4$,
$B=6$, $p=5$, $S\in\{16,64,256\}$), which draws the estimator from the bits
directly rather than through the project's simulators, agrees within $1.48$
standard errors.

\runin{Minibatch sweep on the product family.} The registered comparison of the
shot slope in $B$ with its exact prediction failed on the product family because
79 of 210 cells reached the 32-shot floor of the search grid, which censors the
slope. On the 118 cells that lie above the floor and are predicted to need more
than 45 shots the slope is $-0.79$ \cito{-0.80}{-0.77} against $-0.81$ predicted for
the same cells, an interval that just misses the prediction; the cut is made after
the data were seen, so this is a diagnostic and not a test. A product ladder over
a 32-fold range in $p$ cannot test the minibatch dependence: a budget linear in
$B$ predicts an exponent of $1.990$ there, inside the measured interval too.

\subsection*{Two norms}

Because $\lVert F\rVert_{F}^{2}=\kappa^{2}(\operatorname{tr}F)^{2}$, the leading
risk of Theorem~1 is $(\operatorname{tr}F)^{2}\,pVh/(BSG)$, which does not contain
$\kappa$. Reaching $\eps$ relative to $\operatorname{tr}F$ instead of
$\lVert F\rVert_{F}$ therefore costs $\mathrm{PR}(F)$ times less,
$\BQ^{\mathrm{tr}}\simeq2h\,p^{2}V/(\eps^{2}G)$, and with a fixed cone its exponent
in $p$ stays close to 2 at every minibatch. Over $n=64$ to 256 the budgets of the
registered split-shot sweep give $1.95$, $1.94$, $1.92$ and $1.85$ at $B=12$, 48,
192 and 768 (exact budgets rescaled by $\kappa^{2}$; $1.96$ at every $B$ at leading
order), against $2.12$ to $2.74$ in the Frobenius-relative norm. A fixed-cone
exponent of 2 therefore holds for the trace-relative criterion, and the effective
rank is what the stricter Frobenius-relative criterion adds. Both criteria are
reasonable. The Frobenius-relative error is the plain meaning of ``$F$ to
precision $\eps$'': it holds every direction of $F$ to a fraction of $F$'s size,
so it fails when the shot noise of $p$ coordinates swamps a low-rank $F$. The
trace-relative error suits quantities that depend only on $F/\operatorname{tr}F$,
such as the effective dimension, whose first-order change is
$\delta d\simeq(k/\log k)\operatorname{tr}[(1+k\tilde F)^{-1}\delta\tilde F]$ with
$\tilde F=pF/\operatorname{tr}F$ (\SIeffdim). We give budgets in both norms and
use the Frobenius-relative one by default.

\subsection*{Single-copy optimality of the readout}

For the measured observable the measurement question is settled within
single-copy strategies: $f=n^{-1}\sum_{q}Z_{q}$ is diagonal in the computational
basis, so any single-copy measurement whose estimator is unbiased at every state
satisfies $\sum_{k}x_{k}E_{k}=f$ and hence $\sum_{k}x_{k}^{2}E_{k}\succeq f^{2}$,
and its per-shot variance is at least $\mathrm{Var}_{\rho}(f)$, which the
computational-basis readout attains. Local Pauli shadows\cite{huang2020} pay
exactly $2/n$ more variance per shot than the $Z$ basis on every state, and global
Clifford shadows a variance growing like $2^{n}/n$. Theorem~2 extends the
statement to collective and adaptive measurements.

\subsection*{Proof of Theorem 2}

\emph{One state.} For a state in the full model and $\psi=\operatorname{tr}(\rho
f)$, let $L=f-\langle f\rangle$, $K_{t}=1+\tfrac{t}{2}L$ and
$\rho_{t}=K_{t}\rho K_{t}/\operatorname{tr}(K_{t}\rho K_{t})$. This is a state for
every small $t$ (a congruence of $\rho$, so positive, and pure when $\rho$ is
pure), and since $\operatorname{tr}(\rho L)=0$ its tangent at $t=0$ is
$\tfrac12\{\rho,L\}$. The affine curve $\rho+\tfrac{t}{2}\{\rho,L\}$ has the same
tangent but, for a pure state with $\operatorname{Var}_{\rho}(f)>0$, a negative
eigenvalue at every $t\ne0$, which is why the congruence is used.
The symmetric logarithmic derivative at $t=0$ is $L$, the quantum Fisher
information $\operatorname{tr}(\rho L^{2})=\operatorname{Var}_{\rho}(f)$, and
$\mathrm{d}\psi/\mathrm{d}t=\operatorname{tr}(\rho Lf)=\operatorname{Var}_{\rho}(f)$;
this is the semiparametric bound for an expectation value\cite{tsang2020}.

\emph{Many states, one vector functional.} Take one such curve per configuration,
$t=(t_{c})_{c}$. The model $\bigotimes_{c}\rho_{c}(t_{c})^{\otimes N_{c}}$ has
symmetric logarithmic derivatives acting on different tensor factors, so its SLD
quantum Fisher information is $Q(t)=\operatorname{diag}(N_{c}q_{c}(t_{c}))$, with
each $q_{c}$ continuous and $q_{c}(0)=V_{c}$. Any measurement, adaptive or collective, applied to all copies
at once yields a classical model whose Fisher information matrix obeys
$I(t)\preceq Q(t)$ (Braunstein--Caves\cite{braunstein1994}). The estimand is the
vector $\operatorname{vec}F(\mu(t))$, whose Jacobian $J$ has column $c$ equal to
$V_{c}\,\partial\operatorname{vec}F/\partial\mu_{c}$. The multivariate van Trees
inequality\cite{gill1995}, applied to that classical model with a smooth prior on
the neighbourhood $U_{sN^{-1/2}}$ and $s\to\infty$ after $N\to\infty$, gives for
the quadratic loss $\lVert\cdot\rVert_{F}^{2}$ the local asymptotic minimax bound
$\lim_{s}\liminf_{N}N\sup_{U_{s/\sqrt N}}\mathbb{E}\lVert\hat F-F\rVert_{F}^{2}\ge
N\operatorname{tr}(JQ(0)^{-1}J^{\mathsf T})=\sum_{c}A_{c}V_{c}/\nu_{c}$, which is
Eq.~\eqref{eq:lb}. When the counts are chosen sequentially from earlier readouts,
the chain rule for Fisher information replaces $N_{c}$ in $Q$ by
$\mathbb{E}N_{c}$, and the same bound follows with $\nu_{c}=\mathbb{E}N_{c}/N$; this is the quantum van Trees route of
Ref.~\cite{gillguta2013}. It needs neither full-rank states nor local asymptotic
normality, so it covers the pure configuration states of a noiseless circuit; for
full-rank states with non-degenerate spectra quantum local asymptotic
normality\cite{kahn2009} gives the same bound and its attainability in the
Gaussian limit. A sum of entrywise minimax bounds would not by itself bound the
minimax risk of the sum, because the least favourable direction can differ
between entries, which is why the vector form is needed. The remainder depends
on the number $B(1+2p)$ of configurations, on the $\nu_{c}$ and on the $V_{c}$,
which are polynomial in $n$, but no rate uniform in $n$ is proved, so the theorem
is stated at fixed $n$.

\emph{Attainment.} $f=n^{-1}\sum_{q}Z_{q}$ is diagonal, so the computational-basis
sample mean attains $V_{c}/N_{c}$ for every $c$ at once. Differentiating
$F=B^{-1}\sum_{i}g_{i}g_{i}^{\mathsf T}$ with $g_{i}=2r_{i}d_{i}$ gives
$\partial F/\partial\mu_{i0}=2g_{i}g_{i}^{\mathsf T}/(Br_{i})$ and
$\partial F/\partial\mu_{ij\pm}=\pm r_{i}(e_{j}g_{i}^{\mathsf T}+g_{i}e_{j}^{\mathsf
T})/B$, whose squared norms are the stated $A_{c}$; substituting the leading
covariance $C_{i}$ into $B^{-2}\sum_{i}(2\lVert g_{i}\rVert^{2}\operatorname{tr}C_{i}
+2g_{i}^{\mathsf T}C_{i}g_{i})$ reproduces $\sum_{c}A_{c}V_{c}/N_{c}$ with
$N_{i0}=S_{0}$ and $N_{ij\pm}=S$. \emph{Budget.} Minimising $\sum_{c}N_{c}$ subject
to $\sum_{c}A_{c}V_{c}/N_{c}\le\eps^{2}\lVert F\rVert_{F}^{2}$ gives
$N_{c}\propto\sqrt{A_{c}V_{c}}$ by the Cauchy--Schwarz inequality, hence
Eq.~\eqref{eq:nstar}. In the homogeneous reading the shift terms dominate,
$\sqrt{A_{ij\pm}V}\simeq|r_{i}|\lVert g_{i}\rVert\sqrt{2V}/B$, and the ratio of
uniform to optimal cost is the mean square over the squared mean of
$x_{i}=r_{i}^{2}\lVert d_{i}\rVert$, which is $h$ when $G$ is homogeneous.

\subsection*{The class of Theorem 2 and estimators outside it}

Equidistant $\pm\pi/2$ shifts are already variance-optimal among shift
rules\cite{epsr2025}, and for unbiased estimators of a single gradient from a
shot-noise oracle, a directional forward-gradient rule reaches the
Cram\'er--Rao bound up to a factor $(N-1+\alpha)/N$ that tends to 1 as the number
$N$ of parameters grows, in its bias-neutral limit and under the oracle model of
that work\cite{coyle2026}, which differs from the shift-configuration access
assumed here. Outside the
class are estimators that do not query the shift configurations one parameter
at a time or that use structure: simultaneous perturbation of the quantum Fisher
information\cite{gacon2021}, random natural gradients\cite{kolotouros2024},
Stein-identity estimators\cite{halla2025}, multi-copy schemes\cite{abbas2023},
simultaneous gradient measurement\cite{bowles2025,chinzei2025}, which reads the
whole gradient from a few measurement settings of the unshifted circuit when its
generators commute, and estimators that use a known light
cone\cite{companion}; the companion paper compares the last two on hardware. A bound over
all circuit designs, and over coherent-access algorithms\cite{gilyen2019}, which
beat shot noise but need fault tolerance, is open.

\subsection*{Simulation backends and their validity range}

Exact state vectors are used to $n\approx24$. Above that, the product and blocked
families are simulated exactly as products of single-qubit or $L$-qubit states,
and the brickwork and chain families with a bounded-$\chi$ matrix-product-state
backend (relative truncation $10^{-8}$, bond-dimension cap 64), validated against
exact state vectors to the double-precision floor and with an exact-identity gauge
null for each gradient component (\SIfixed). For the blocked circuit each block
is simulated as a $2^{L}$-dimensional state vector, and a joint shot is sampled
exactly as independent draws of each block's sum of $Z$ outcomes; the
construction reproduces a full state-vector simulation of the same circuit to
$10^{-16}$ at $n=8$, and its Monte Carlo error matches Eq.~\eqref{eq:exact}
($z=0.08$).

\subsection*{Matched-precision protocol}

Two kinds of cost number appear in the main text, and each result says which.
\emph{Measured} costs come from a matched-precision shot search: the smallest $S$
on a geometric grid (ratio $1.3$ for the 256-qubit product ladder, $1.4$ elsewhere) at which the median
relative Frobenius error of three independent estimates is at most $\eps$, against
the exact Fisher matrix of the same samples; $\BQ=B(1+2p)S$.
\emph{Theorem-1 predictions} solve Eq.~\eqref{eq:exact} for $S$ at relative
mean-square error $\eps^{2}$, or its leading order Eq.~\eqref{eq:cost} where
stated; nothing is fitted. A third kind, fixed-constant predictions that hold
$h/(B\kappa^{2})$ constant, is used only in \SIfixed. $\eps$
is $0.15$ unless stated.

\subsection*{Hardware protocol}

\runin{Line runs.} The model is a six-qubit chain with depth 2 and
$p=12$, trained noiselessly with a seed fixed in the plan and evaluated on $B=8$
binary inputs, run on a physical line of
\device{ibm\_marrakesh} chosen from live calibration (minimum summed two-qubit and
readout error; the line is stored in each record). Each job executes all
$B(1+2p)=200$ configurations, transpiled once at optimisation level 1, with
dynamical decoupling and twirling off. The uniform job gives each configuration
2,048 shots; the allocated job distributes the same 409,600 shots by minimising
the exact risk at the first job's single-shot moments, with every configuration
receiving at least 64 shots. Each job is cut into 32 contiguous replicate blocks
of the returned shots, each block giving one split-shot estimate
$\hat F_{r}$; the observed error is
the mean of $\lVert\hat F_{r}-\bar F_{-r}\rVert_{F}^{2}$, which estimates the
risk plus the leave-one-out variance of the reference, and the prediction is
computed for the same quantity. The confidence radius is computed per block
from that block's own moments. Intervals bootstrap the 32 blocks. The registered
gates are: the interval of observed over predicted for the uniform job inside
$[0.67,1.5]$; the interval of the allocation's error reduction above 1 and
containing the prediction from the first job; and coverage of at least 55 of 64.

\runin{Capacity-measure runs.} The product run evaluates a trained 16-qubit model
without entangling gates ($p=32$, $B=64$) as eight copies on 128 qubits of
\device{ibm\_kingston}, each copy on different inputs, with 4,096 shots for each
of its 520 circuits; subsets of the shots give the estimates at fewer shots. The
entangling run evaluates the depth-2 CNOT brickwork on a line of 48 qubits
($p=96$, $B=128$) as two copies on two 48-qubit lines of the same processor. Its
gradients are measured with the light-cone design of the companion
paper\cite{companion}: the parameters are coloured so that the light cones of the
members of a class, the qubits their gates can reach, lie apart, every class is
shifted by $\pm\pi/2$ in one circuit, and each member's derivative is read as the
mean $Z$ of the qubits of its own light cone that its gate actually influences. If
the device respects the light cones, each derivative estimate is an unbiased shift
difference with a variance that a pilot measures, and the standard estimate, the
split-shot estimate and Theorem~1's exact risk apply with the statistics of the
qubits read, parameters shifted in the same circuit being treated as independent.
The run's direct test, which repeats a subset of derivatives with the parameter
shifted alone, bounds the departure (Results); the companion paper analyses it in
full. Counts
are stored per 32-shot chunk, and the bootstrap of the bracket and of the corrected
estimate resamples each input's chunks. The run's other arms, used only in the
companion paper, share its jobs.

\runin{Comparison at six parameter points.} The depth-2 CNOT brickwork on a line of
16 qubits ($p=32$, $B=64$ binary inputs, parity labels, $f$ the mean $Z$ of all 16
qubits) runs as six copies on six disjoint 16-qubit lines of
\device{ibm\_marrakesh}, copy $c$ carrying parameter point $c$: three uniform draws
and three points trained noiselessly by 100 Adam steps on the squared loss. Every
circuit runs one input at one configuration (unshifted, or one parameter shifted
by $\pm\pi/2$) on all six copies, $4{,}160$ circuits at 2,048 shots each, with
dynamical decoupling, twirling off and no readout mitigation; counts are stored
per 32-shot chunk. Standard estimates at 256 to 2,048 shots come from 20 random
subsets of the chunks, the split-shot estimate at all shots from 20 random
halvings, and the jackknife from 16 groups of four chunks. The classical networks
have one hidden layer of two tanh units (37 parameters); their matrices are exact.
The plan was registered on another processor of the same family and moved to this
one before any data, when that processor went into maintenance; lines and
predictions were redone on the new device by the registered rules. Predictions
come from 20 replicates that draw Gaussian chunk means with the exact means and
covariances of the 16 outcomes, after the frozen readout errors.

\runin{Bracket for the effective dimension.} Let $\hat F$ be the split-shot
estimate at all shots, not clipped, with eigenvalues $\mu_{m}$ and trace $t$. A
bootstrap that resamples each input's chunks gives draws of the split-shot
error about the standard estimate; $r$ is the $97.5\%$ quantile of their largest
eigenvalue in absolute value and $r_{t}$ that of their trace in absolute value.
If $\lVert\hat F-F\rVert_{\mathrm{op}}\le r$ and $|\operatorname{tr}\hat
F-\operatorname{tr}F|\le r_{t}$, Weyl's inequality puts each eigenvalue of $F$
within $r$ of $\mu_{m}$, and since $d$ increases with each eigenvalue and
decreases with the trace,
\begin{equation}
  \sum_{m}\frac{\log\bigl(1+kp(\mu_{m}-r)_{+}/(t+r_{t})\bigr)}{\log k}
  \le d(F)\le
  \sum_{m}\frac{\log\bigl(1+kp(\mu_{m}+r)/(t-r_{t})\bigr)}{\log k}.
  \label{eq:dbracket}
\end{equation}
The bracket is asymptotic, like the confidence radius. Its half-width is not
proportional to $r$, because the lower end sets every eigenvalue below $r$ to zero;
the shots needed for a stated half-width follow from evaluating
Eq.~\eqref{eq:dbracket} with $r$ and $r_{t}$ scaled as $S^{-1/2}$, which is how the
multiple quoted in Results was computed.
In 40 simulated runs of the brickwork design under a readout-noise model the
bracket contained the true $d$ 40 times, and the operator-norm event held 38
times (nominal 39).

\subsection*{Statistics, seeds and reproducibility}

All intervals are two-sided 95\% and come from 3000-resample bootstraps. Where
replicates share a device condition or a random seed, the bootstrap resamples the
clustering unit (design points or replicate blocks for hardware, seeds for
simulation) and not the rows. Exponents are ordinary least squares on logarithms
with the seed-clustered interval; where families with different intercepts are
pooled, a fixed-effects specification with one intercept per family is used. No
$P$ values are computed; all tests are interval-containment tests against
predicted values declared
before the corresponding cells were generated. Each experiment draws its design
(parameters, inputs and labels) from a generator seeded by a value fixed in its
plan, and its shots from a separate generator keyed by seed, minibatch and
estimator, so that estimators compared on one cell share the design and differ
only in their shots; all seeds are stored in the records. Sample sizes are given
with each result.

\subsection*{Registration of predictions}

Each registered experiment has a plan and script committed to the project
repository before it ran, recording the question, the prediction, the pass and
failure conditions and the analysis; there is no entry in an external registry.
Failures are reported against the condition as written, and a re-run after a
failure has its own plan. A manifest of the hash of every plan was later
timestamped publicly with OpenTimestamps, anchored in Bitcoin; it shows that no
plan has changed since, while the dates before the runs rest on the commit
history. Table~\ref{tab:registration} gives the status of every
experiment the main text cites; \SIfailures{} gives each failed check with its
prediction, measurement and consequence.

\begin{table}[tbp]
\centering\footnotesize
\caption{\textbf{Registration status of the experiments cited in the main text.}
Registered: plan and script committed before the run. Exploratory: plan hashed
before the run but committed afterwards, or no plan. A registered experiment
whose checks did not all clear is listed with its failures in \SIfailures.}
\label{tab:registration}
\setlength{\tabcolsep}{5pt}
\begin{regtable}
\toprule
\textbf{Experiment} & \textbf{Status} & \textbf{Registered checks}\\
\midrule
Theorem~1 checks (exact against Monte Carlo) & exploratory & none\\
Minibatch sweep, $B=4$ to 256 & exploratory & product shot-slope check failed (grid floor)\\
Product family to 256 qubits & registered & precision-exponent check failed narrowly\\
Blocked circuits at $B=12$, $n\le128$ & exploratory & regime-difference check failed\\
Effective-rank sweep, standard estimator & registered & rise and $B=768$ exponent failed; two others pass\\
Effective-rank sweep, split-shot estimator & registered & rise and two others pass; $B=768$ exponent and constant heterogeneity factor failed\\
Bounds, allocation, confidence radius, effective dimension & exploratory & none\\
Trained models on real data & registered & all three failed\\
Hardware estimate of $F$, first run & exploratory & all pass\\
Hardware estimate of $F$, replication & registered & all pass\\
Readout weights on hardware & registered & four pass; weight saving failed\\
Effective dimension on hardware, product model & registered & all three pass\\
Effective dimension of an entangling model on hardware & registered & the three of this paper pass\\
Capacity comparison with classical networks & exploratory, after the run & none\\
Capacity comparison at six parameter points & registered & all three pass\\
\bottomrule
\end{regtable}
\end{table}

\subsection*{Use of large language models}

A large language model (Claude, Anthropic) was used under the authors'
direction for code, analysis scripts, drafting and editing of text, and checking
derivations. The authors take responsibility for the content and for every number
reported. The model is not an author.

\section*{Data availability}

The datasets generated and analysed in this study comprise every locked JSON
evidence record, including all hardware execution results, and are included in
the reproduction archive described under Code availability. Each numerical claim
in this manuscript binds to a named field in one of those records; the mapping
is given in \SIclaims.

\section*{Code availability}

The \device{shotcost} budget tool, the simulation backends, the experiment
scripts, the registration plans and the gate checker are in the reproduction
archive, which rebuilds every figure from the records. A copy is provided to the
editor and referees with the submission, and the archive will be deposited on
Zenodo, with a DOI, before publication.

\section*{Acknowledgements}

This study received no funding. The cost law at matched precision and an earlier transfer study were presented at two non-archival NeurIPS 2026 workshops (SaTQuML and Sim2Science).

\section*{Author contributions}

P.S. designed the study, derived the theorems, implemented the simulation
backends and the experiment pipeline, executed the hardware runs, performed the
statistical analysis, and wrote the manuscript. C.L. contributed to the
experimental design, the analysis, and the revision of the manuscript. Both
authors read and approved the final manuscript.

\section*{Competing interests}

The authors declare no competing financial or non-financial interests.



\clearpage
\appendix
\renewcommand{\thefigure}{A\arabic{figure}}
\renewcommand{\thetable}{A\arabic{table}}
\setcounter{figure}{0}
\setcounter{table}{0}


\section{Registered checks that did not clear}
\label{app:failures}

Table~\ref{tab:failures} lists every check registered in advance that did not
clear and that bears on the results of this version, and what each one changed.
Several are about the instrument rather than the physics: a pooled estimator, a
point null on a derived quantity, and two shot searches whose grid was coarser
than the interval they were asked to resolve. They are recorded against the
condition as written. Registered checks on material of an earlier version, and on the light-cone
measurement of the companion paper, are recorded in the reproduction archive
and in that paper.

\paragraph{The effective-rank sweeps.} The main text states that four registered
checks of the two effective-rank sweeps failed. The split-shot sweep asked the
$B=768$ exponent over $n=64$ to 256 to contain 3 with a lower bound of at least
$2.75$; the measured $2.74$ \ci{2.69}{2.79} fails by $0.06$. The miss sits in the
trace-relative part of the budget and not in the effective rank. Each
Frobenius-relative exponent is the trace-relative one minus the slope of
$\kappa^{2}$: $1.95+0.14$, $1.94+0.39$, $1.92+0.71$ and $1.85+0.90$ give $2.09$,
$2.33$, $2.63$ and $2.75$, against $2.12$, $2.36$, $2.64$ and $2.74$ measured. At
$B=768$ the trace-relative exponent is $1.85$ against $1.96$ at leading order, the
finite-shot part of the exact risk, so the premise that the exponent approaches 3
at $B=768$ and $n\le256$ does not hold. A second check asked the ratio of the
exact-risk budget to $p^{2}V/(\eps^{2}\kappa^{2}G)$ to vary by less than 1 across
cells, and fails ($2.03$ to $3.77$). That ratio was registered as the
heterogeneity factor, but it is $2h$ times a finite-shot factor, so it says that
one $h$ does not fit all cells; the budgets are therefore quoted from the exact
risk, which carries $h$ cell by cell. The first sweep, with the standard
estimator, failed its rise and $B=768$ checks because that estimator's finite-shot
bias flattens the exponent at large $B$. A cost exponent near 3 at $B=12$ is a
property of a light cone that grows with the register; with a fixed cone the
exponent lies between 2 and 3 and is set by the effective rank, which the
minibatch controls.

\paragraph{The blocked arms at $B=12$.} That run was exploratory, and its
registered contrast asked for the difference between the growing-cone and the
fixed-cone exponents to exceed $0.5$ at its lower bound; the difference is $0.76$
\ci{0.45}{1.01}.

{\footnotesize
\setlength{\tabcolsep}{5pt}
\setlength{\LTcapwidth}{\textwidth}
\begin{longtable}{@{}>{\raggedright\arraybackslash}p{3.4cm}>{\raggedright\arraybackslash}p{2.9cm}>{\raggedright\arraybackslash}p{3.1cm}>{\raggedright\arraybackslash}p{4.5cm}@{}}
\caption{\textbf{Registered checks that did not clear.} Each row gives the
prediction as registered before the corresponding cells were generated, the
measured value, and what the failure changed in the paper. Intervals are
two-sided 95\%. $\dagger$~Plan hashed and stored before the run but committed to
the repository after it; reported as exploratory.}
\label{tab:failures}\\
\toprule
\textbf{Check} & \textbf{Predicted} & \textbf{Measured} & \textbf{Consequence}\\
\midrule
\endfirsthead
\toprule
\textbf{Check} & \textbf{Predicted} & \textbf{Measured} & \textbf{Consequence}\\
\midrule
\endhead
\bottomrule
\endlastfoot
$\eps$-exponent (256-qubit product ladder) & $-2$ & $-1.92$ $(-1.995,-1.845)$ &
Attributed to shot-grid granularity; reported as a partial miss\\
Brickwork at scale & small-$n$ CI overlaps $n\ge64$ &
$2.853\rightarrow1.715$, disjoint &
Cubic exponent does not survive size on this family\\
Blocked-cone sweep, pooled estimator & pooled exponent $=2$ &
$2.513$ pooled, $1.949$ fixed-effects &
Estimator defect; re-run rather than re-scored\\
Blocked-cone sweep, fixed effects & CI contains 2 and 3 exactly &
$1.984$ and $3.034$ &
Point null on a derived quantity; no third attempt\\
Minibatch sweep, product family$^{\dagger}$ & shot slope in $B$ equals exact
prediction ($-0.67$) & $-0.50$ $(-0.52,-0.48)$ &
79 of 210 cells at the 32-shot grid floor, which censors the slope; uncensored
cells give $-0.79$ against $-0.81$; chain passes\\
Blocked arms at $B=12$, regime difference$^{\dagger}$ & CI lower bound $>0.5$ &
$0.76$ $(0.45,1.01)$ &
Regimes separate by less than the fixed-constant prediction ($1.10$);
fixed-cone exponent reported as $2.27$ at $B=12$\\
Effective-rank sweep, standard estimator & rise of the fixed-cone
exponent from $B=12$ to 768, CI lower bound $>0.3$ & $0.16$ $(0.10,0.23)$ &
The standard estimator's finite-shot bias flattens the exponent at large $B$;
re-run with the split-shot estimator\\
Effective-rank sweep, standard estimator & $B=768$ exponent over
$n\ge64$: CI contains 3, lower bound $\ge2.75$ & $2.56$ $(2.51,2.61)$ &
As above\\
Effective-rank sweep, split-shot estimator & $B=768$ exponent over $n\ge64$:
CI contains 3, lower bound $\ge2.75$ & $2.74$ $(2.69,2.79)$ &
The shortfall sits in the trace-relative part ($1.85$ against $1.96$ at leading
order, the finite-shot term); the exponent is given as a function of $B$\\
Effective-rank sweep, split-shot estimator & budget over
$p^{2}V/(\eps^{2}\kappa^{2}G)$, registered as $h$, varies by $<1$ & $2.03$ to $3.77$ &
That ratio is $2h$ times a finite-shot factor; budgets quoted from the exact
risk, which carries $h$ cell by cell\\
Trained models, $B$ through PR & budget per unit PR, $B=16\to256$,
median in $[0.67,1.5]$ & $1.56$ (range $1.20$--$1.84$) &
Leading order gives $1.02$; the rise is the finite-shot part of the exact risk\\
Trained models, PR as the price of the norm & median of
$(N_{F}/N_{\mathrm{tr}})/\mathrm{PR}$ in $[0.9,1.1]$ & $0.83$ (range $0.58$--$0.97$) &
Exactly 1 at leading order; the exact price is stated as $0.7$ to $0.9$ of PR
at these budgets\\
Trained models, training lowers PR & median ratio $<1$, CI upper
bound $<1$ & $0.90$ $(0.81,1.05)$ &
No claim that training lowers the effective rank; it lowers $V/G$\\
Readout weights on hardware & weight saving CI lower bound $>1$ &
$0.68$ $(0.48,0.98)$ &
Certification-optimal readout weights are not pursued\\
\end{longtable}}

\section{Cost at scale, the bound in practice, and fixed-constant predictions}
\label{app:fixed}

\paragraph{What this part adds.} The main text states the cost law and its check on
a product family; this part gives the sweeps behind its statements about the
effective rank and the cost exponents on blocked circuits, with the registered
checks that failed. Equation numbers of the main text are cited as numbers.

\paragraph{The effective rank on blocked circuits.}
The prefactor $V/G$ of Eq.~(5) is set by how far each gate's influence
spreads, its light cone. If every Heisenberg-evolved $Z_{q}$ is supported on at
most $w$ qubits, then $\operatorname{Cov}(Z_{q},Z_{q'})=0$ for disjoint supports
and $\partial_{j}\langle Z_{q}\rangle=0$ unless qubit $q$ lies in the light cone of
gate $j$, so $nV$ and $nG$ are bounded independently of the register size and
$V/G$ stays fixed as qubits are added (eleven families below). The effective
rank is not fixed by the circuit. On a blocked circuit, which applies a chain of
CNOT gates inside disjoint blocks of $L$ qubits and never across them (Methods),
the overlap of two samples' gradients is a sum of $n/L$ block terms that are
independent under independently drawn parameters, so
\begin{equation}
  \kappa^{2}\approx\frac{a}{B}+c\,\frac{L}{n},
  \label{eq:kappablocks}
\end{equation}
the first term from the $B$ diagonal entries of
$\lVert F\rVert_{F}^{2}=B^{-2}\sum_{i,k}(g_{i}\cdot g_{k})^{2}$ and the second
from a random walk of $n/L$ steps in the off-diagonal ones. This is a
phenomenological form, not a derived law. Fitted over the 80 fixed-cone cells of
a registered sweep ($n=16$ to 256, $B=12$ to 768), $a=1.06$, $c=0.44$,
$R^{2}=0.96$ (Fig.~\ref{fig:effrank}a). For $n\ll cLB$ the effective rank grows
like $n/(cL)$ and the exponent of the budget in $p$ approaches 3; for $n\gg cLB$ it
returns to 2. A cost exponent fitted to one family is therefore a property of the
data and the minibatch as well as of the circuit.

\begin{figure}[tbp]
\centering
\includegraphics[width=\textwidth]{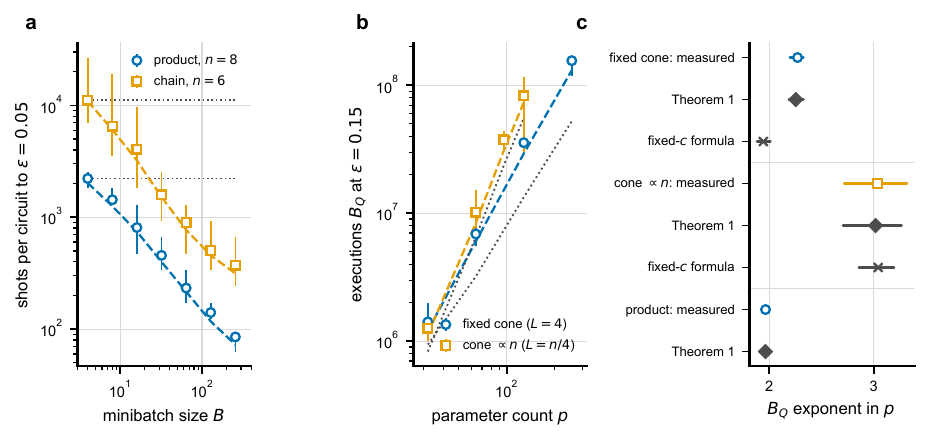}
\caption{\textbf{The exact risk predicts measured cost.}
\textbf{a}~Shots per circuit needed to reach $\eps=0.05$ against minibatch
size $B$, from a matched-precision shot search on the product family ($n=8$, blue
circles) and the chain ($n=6$, orange squares). Markers are geometric means over
10 seeds and bars span the seeds. Dashed lines are the prediction of
Theorem~1 from exact moments; dotted lines show the constant shot count that a
budget growing linearly with $B$ would imply.
\textbf{b}~Measured executions $\BQ$ to $\eps=0.15$ at $B=12$ against parameter
count on the blocked circuit, with the light cone fixed at $L=4$ qubits (blue
circles) or growing as $L=n/4$ (orange squares), for $n=16$ to 128. Markers are
geometric means over 6 seeds, bars span the seeds, dashed lines are the
prediction of Theorem~1 from exact moments, and grey dotted lines are the
fixed-constant shortcut below, which holds $\kappa$ fixed.
\textbf{c}~$\BQ$ exponents in $p$ with two-sided 95\% seed-clustered bootstrap
intervals: measured (open markers), predicted by Theorem~1 (diamonds) and
predicted with a fixed constant or by a budget linear in $B$ (crosses), for the
two blocked arms and the product family to 256 qubits.}
\label{fig:validation}
\end{figure}

\begin{figure}[tbp]
\centering
\includegraphics[width=\textwidth]{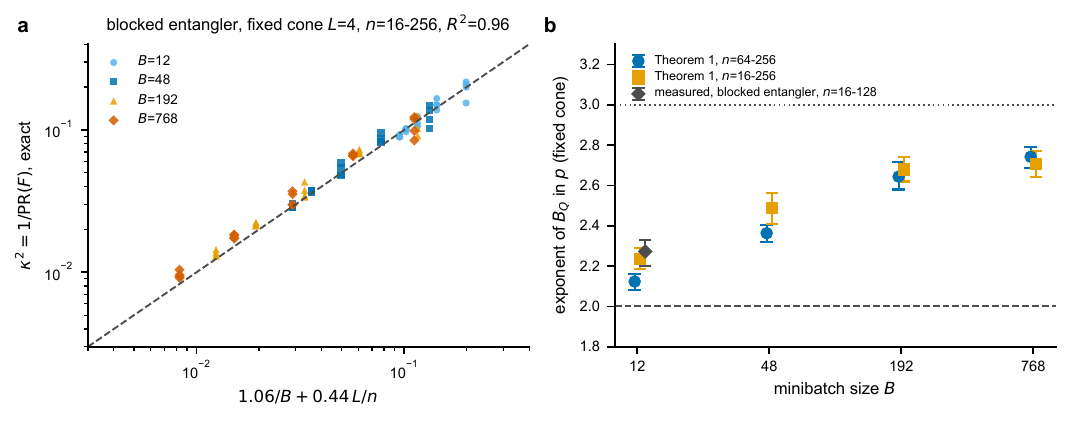}
\caption{\textbf{The effective rank sets the exponent when the light cone is
fixed.} \textbf{a}~Exact $\kappa^{2}=1/\mathrm{PR}(F)$ against the fit
$a/B+c\,L/n$ of Eq.~\eqref{eq:kappablocks}, for the 80 fixed-cone cells of the
registered effective-rank sweep with the standard estimator (blocked circuit,
$L=4$, depth 2, $n=16$ to 256, five seeds), coloured by minibatch size; the dashed
line is equality. \textbf{b}~Exponent of the exact-risk budget $\BQ$ in $p$ with
the cone fixed, against $B$, from the registered sweep with the split-shot
estimator over $n=64$ to 256 (blue circles) and $n=16$ to 256 (orange squares),
with two-sided 95\% seed-clustered bootstrap intervals; the grey diamond is the
measured shot-search exponent at $B=12$ over $n=16$ to 128. Dashed and dotted
lines mark 2 and 3: the exponent rises from $2.12$ to $2.74$ as $B$ grows.}
\label{fig:effrank}
\end{figure}

\paragraph{Measured cost against the exact risk.}
The exact mean-square error agrees with 4,000-repetition Monte Carlo in all 12
cells tested (no deviation beyond $1.29$ standard errors), and the split-shot
estimator with its own expression in all four (largest $|z|=1.31$). The
leading-order expression of Eq.~(2) is within $0.6\%$ of the exact value at
$S=1024$ and up to $26\%$ low at $S\le64$. A matched-precision shot search over
$B$ from 4 to 256 (Fig.~\ref{fig:validation}a; exploratory) gives slopes of
$\log\BQ$ on $\log B$ of $0.50$ \ci{0.48}{0.52} on a product family and $0.30$
\ci{0.25}{0.35} on a CNOT chain, both excluding the linear growth that a
triangle-inequality bound gives; on the chain the shot count's slope, $-0.70$
\cito{-0.75}{-0.65}, matches the prediction of $-0.70$, while on the product family
the comparison failed because 79 of 210 cells reached the floor of the search grid.
On a product family to 256 qubits (registered) Theorem~1 predicts an executions
exponent in $p$ of $1.965$, against $1.966$ \ci{1.934}{1.997} measured.

Blocked circuits, exact products of $L$-qubit blocks, make a genuine shot search
affordable at 128 qubits. At $B=12$ the measured $\BQ$ exponent is $2.27$
\ci{2.20}{2.33} with the cone fixed at $L=4$ and $3.04$ \ci{2.72}{3.32} with the
cone growing as $L=n/4$, against $2.26$ and $3.02$ predicted
(Fig.~\ref{fig:validation}b,c; exploratory, and its registered contrast failed,
\SIfailures); the median ratio of measured to predicted shots, $1.18$, is the
expected overshoot of a grid spaced by $1.4$. The fixed-cone exponent exceeds 2
because $\kappa$ falls from $0.46$ to $0.32$ as blocks are added, which by the
Lemma is the effective rank. In two registered sweeps over $n=64$ to 256 the
fixed-cone exponent of the split-shot budget rose from $2.12$ \ci{2.08}{2.16} at
$B=12$ through $2.36$ and $2.64$ to $2.74$ \ci{2.69}{2.79} at $B=768$
(Fig.~\ref{fig:effrank}b); the registered rise, $0.47$ \ci{0.41}{0.53}, cleared its
bar of $0.3$, and four checks failed, among them an exponent of 3 at $B=768$, short
by $0.06$ (\SIfailures).

The choice of yardstick decides whether the effective rank enters at all. Measured
relative to $\operatorname{tr}F$ it drops out,
$\BQ^{\mathrm{tr}}\simeq2h\,p^{2}V/(\eps^{2}G)$, and the same budgets give
exponents of $1.95$ to $1.85$. The extra power of $p$ is therefore the price of
resolving every direction of a matrix of low effective rank to a fraction of its
Frobenius norm; it is not a property of the measurement. Which yardstick applies
depends on the use: quantities of $F/\operatorname{tr}F$ such as the effective
dimension need only the trace-relative error, while a preconditioner or a
statement about individual directions needs the Frobenius-relative one (Methods
gives budgets in both). In either case the shots for a stated precision follow
from Theorem~1 before the run, and Eq.~\eqref{eq:kappablocks}, a fit, says when
the effective rank grows with the register.

\begin{figure}[htbp]
\centering
\includegraphics[width=\textwidth]{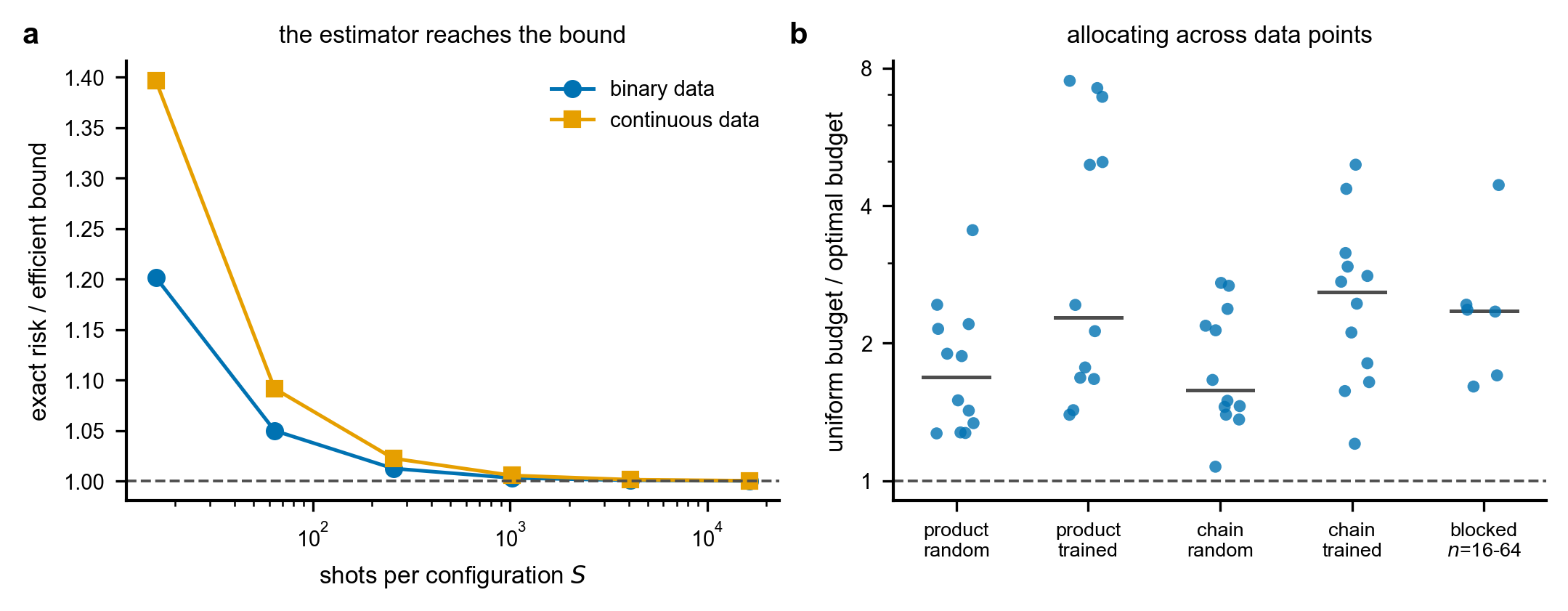}
\caption{\textbf{The bound is attained, and allocation across data points pays.}
\textbf{a}~Exact risk of the split-shot estimator with uniform shots over
the lower bound of Theorem~2, against shots per configuration, for binary (blue
circles) and continuous (orange squares) data on a product model.
\textbf{b}~Executions under uniform shots over executions under the exact-risk
allocation, per cell (points) and median (bar), for product and chain models at
random and trained parameters ($\eps=0.1$) and the blocked circuit at $n=16$ to
64 ($\eps=0.1$, $B=12$). All exploratory.}
\label{fig:tightness}
\end{figure}

\paragraph{The bound is attained, and allocation across data points pays.}
Figure~\ref{fig:tightness} gives the two results of the main text on Theorem~2 in
practice: the split-shot estimator's exact risk approaches the bound as the shots
per configuration grow ($1.20$ at $S=16$ to $1.0002$ at 16,384), and allocating
shots across samples by the exact risk saves $1.6$ to $2.6\times$ in median per
family ($1.1$ to $7.5\times$ across cells), with the blocked circuit at $n=16$ to 64 needing $1.61$ to $4.44$ times
fewer executions. All of these are exploratory.

\begin{figure}[htbp]
\centering
\includegraphics[width=\textwidth]{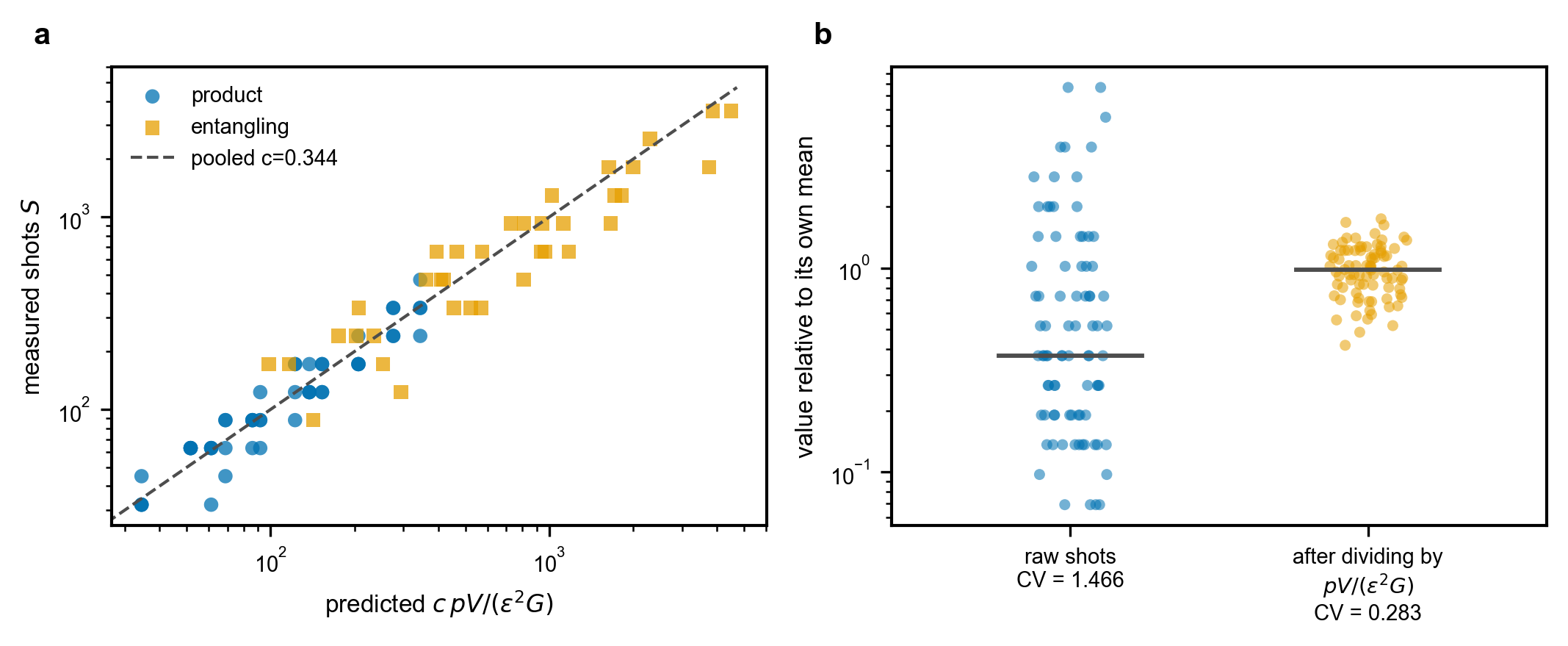}
\caption{\textbf{At fixed minibatch size one constant reproduces the cost of
both families.}
\textbf{a}~Measured shots against $pV/(\eps^{2}G)$ on logarithmic
axes for 84 cells at $\eps=0.15$ and $B=12$, product family as blue circles and
entangling family as orange squares; the grey dashed line is the pooled constant
$0.344$ fitted across both families and not per family. \textbf{b}~The 84
per-cell values before dividing by $pV/(\eps^{2}G)$ (blue, left) and after
(orange, right), each cell drawn as an individual point and each column
normalised to its own mean, on a logarithmic axis; the grey horizontal bar is the
column median and the annotation is the pooled coefficient of variation over all
84 cells.}
\label{fig:constant}
\end{figure}

\paragraph{One constant at fixed minibatch.} At fixed $B$ one constant absorbs
$h/(B\kappa^{2})$. At $B=12$ the measured ratio $S/(pV/(\eps^{2}G))$ is $0.338$
for an entangling chain and $0.349$ for the product family, a 3\% difference, and
the pooled coefficient of variation of that ratio is $0.283$, down from $1.466$
on the raw shot counts (Fig.~\ref{fig:constant}); a third family, a brickwork
circuit, gives $0.341$ \ci{0.306}{0.376} at the same $B$. The same families
have $\BQ$ exponents of $2.856$ \ci{2.363}{3.331} and $1.990$ \ci{1.814}{2.174},
and an independent measurement on the chain with a bit-flip encoding and $B=64$
gives $3.087$ \ci{2.645}{3.529}. The constant is not universal: in the minibatch
sweep on the chain it falls from $0.54$ at $B=4$ to $0.04$ at $B=256$, as
Theorem~1 requires. This is also where the first registered check failed. The
two-family constant experiment predicted $V=\Theta(1)$ for the entangling family;
measured $\dlog{nV}/\dlogn$ is $0.37$ against a predicted 1, so $V$ decays as
$n^{-0.63}$. The constant survives, because it divides out the measured $V$; what
the failure removes is a readout-correlation reading of the cost regimes.

\begin{figure}[htbp]
\centering
\includegraphics[width=\textwidth]{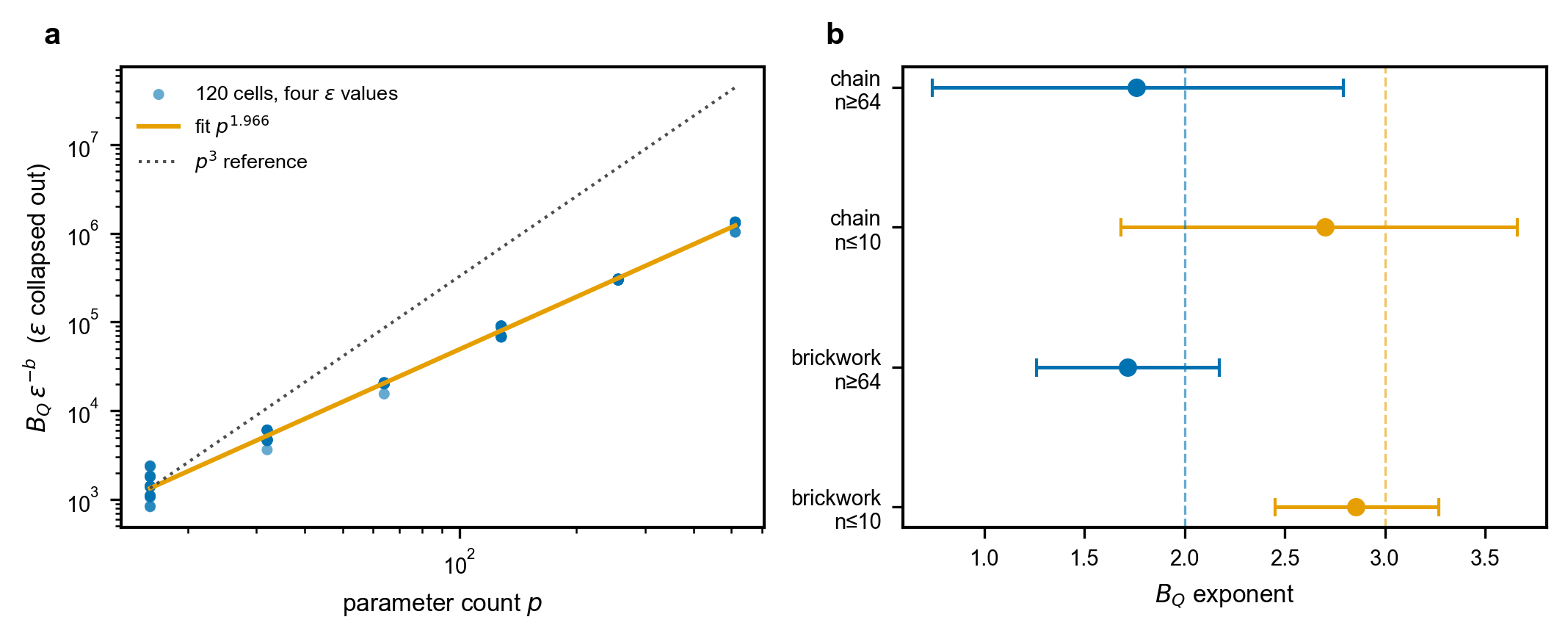}
\caption{\textbf{Fixed-constant predictions.} \textbf{a}~Measured
product-family executions with the fitted $\eps$ dependence divided out:
$\BQ\eps^{-b}$ against parameter count $p$ on logarithmic axes, $n=16$ to 256,
where $b=-1.92$ is the fitted $\eps$ exponent. Blue points are 120 cells at four
$\eps$ values. The orange line is the two-variable fit in $p$; the grey dotted
line is a $p^{3}$ reference sharing the left-hand intercept. \textbf{b}~$\BQ$
exponents predicted from exactly computed $V$ and $G$ with a fixed constant, with
two-sided 95\% seed-clustered bootstrap intervals from 3,000 resamples, for the
registered small-$n$ reference (orange markers) and the $n\ge64$ evaluation (blue
markers) of each entangling family. The dashed lines mark 2 and 3. On the
blocked circuit the same shortcut underestimates the measured fixed-cone
exponent by $0.33$ (main-text Fig.~3c).}
\label{fig:window}
\end{figure}

\paragraph{The product family to 256 qubits.} On the product family, which is
exactly solvable and carries no entanglement, we estimate $F$ to $n=256$ over a
32-fold range in $p$ from 16 to 512 on a geometric shot grid of 120 cells at
$B=16$. The shot exponent is $0.974$ \ci{0.943}{1.005} with $R^{2}=0.996$. The
executions exponent is $1.966$ \ci{1.934}{1.997} (Fig.~\ref{fig:window}a),
against $1.965$ from Theorem~1 on the same cells; collapsing by $\BQ\eps^{2}/p^{2}$
cuts the coefficient of variation from $2.16$ to $0.19$. Budgeting $p^{3}$ from
the small end of that range overestimates the 256-qubit endpoint by a factor of
about 36. One registered sub-check did not clear: the $\eps$-exponent is $-1.92$
($-1.995$ to $-1.845$), missing the predicted $-2$ by $0.005$ at the interval
edge. Theorem~1 on the same cells predicts $-1.99$, so the bias floor does not
explain it; we attribute it to the 30\% granularity of the shot grid and the
median criterion of the search.

\paragraph{Brickwork and chain circuits.} For entanglers whose cone cannot be
fixed by construction we have no measured cost at scale, only the prediction
$S=\lceil c\,pV/(\eps^{2}G)\rceil$ with $c=0.338$ and $V$ and $G$ computed exactly
at one input per cell, using a bounded-$\chi$ matrix-product-state backend. On the
blocked circuit this shortcut reproduces the growing-cone exponent ($3.04$) and
underestimates the fixed-cone one by $0.33$ ($1.95$ against $2.27$), because it
holds $\kappa$ fixed, so the numbers below carry that known downward bias. All
are at depth 2 on a one-dimensional coupling topology, and two families were
registered with their small-$n$ reference locked before any large-$n$ cell was
generated. On the brickwork circuit the predicted exponent is $2.853$
\ci{2.453}{3.268} at $n\in\{4,6,8,10\}$ and $1.715$ \ci{1.262}{2.172} at
$n\in\{64,80,96\}$; the intervals do not overlap (Fig.~\ref{fig:window}b), and the
registered gate, which asked for overlap, fails. On the sequential chain the
locked $n\le10$ interval is $2.700$ \ci{1.681}{3.661}; over $n=16$ to 96 the
predicted exponent is $3.032$ \ci{2.803}{3.245}, and restricted to $n\ge64$ it is
$1.758$ \ci{0.739}{2.794}. The chain's registered overlap test passes without
discriminating, because the locked interval contains both 2 and 3; we record
beside the pass, as a declared non-gating diagnostic, that the $n\ge64$ interval
excludes 3.

Along a sequential chain the gradient attenuates with distance from the readout,
so the number of parameters that measurably move the output stops growing with
the register: over a stratified sample of parameter indices that count is $32$,
$64$, $94.5$, $102.5$, $91.5$ and $96.5$ at $n=16,32,48,64,96,128$, while $p=2n$
grows to 256. The registered positional test fails narrowly: the attenuation rate
is $n$-independent to a relative slope spread of $0.2083$ against a declared
threshold of $0.20$. A survey of eight configurations across product, brickwork
at three depths, chain, ring, and two cat-seeded variants finds no family whose
$nV$ grows with $n$.

\paragraph{What the fixed-constant predictions contain.} Because they hold the
constant fixed, their exponent in $p$ decomposes by construction as
$2+\dlog{nV}/\dlogn-\dlog{nG}/\dlogn$ plus the finite-$p$ contribution of the
$(1+2p)$ prefactor, which a 624-cell sweep over eleven blocked families
reproduces to within $0.001$. That agreement is algebra and carries no evidence of
its own: the content of the sweep is the scaling of $nV$ and $nG$ that it computes
exactly, with $nG$ flat when the cone is fixed and falling as $1/n$ when it is
proportional. Two registered gates asked whether this
derived exponent equals exactly 2 and 3; both failed (Table~\ref{tab:failures}).

\paragraph{Matrix-product-state validity.} Simulator validity is checked at the
level of the quantity being measured. A componentwise comparison against exact
state vectors at every $n$ where that is affordable gives $\max_{j}\lvert
g_{j}^{\mathrm{MPS}}-g_{j}^{\mathrm{SV}}\rvert\approx1.5\times10^{-16}$. A
bond-dimension pair below the realised $\chi$ must move $G$ and one above it must
not; a pair in which both caps exceed the realised $\chi$ cannot bind and is
recorded as vacuous. An exact-identity gauge null, in which every rotation angle
is shifted by $2\pi$, gives the roundoff floor of each gradient component
directly. That check excludes the ring family at $n\ge80$, where the mean-$Z$
expectation falls to $10^{-16}$ and the gauge null perturbs each gradient
component by more than its own value; the families reported are above that floor
by at least seven orders of magnitude.

\paragraph{A readout in two bases.} For $f=n^{-1}\sum_{r}(\cos\phi\,Z_{r}+\sin\phi\,X_{r})$,
$\phi=\pi/3$, each configuration is run twice, read in the computational basis
for the $Z$ part and after a Hadamard on every qubit for the $X$ part, with a
fraction $\pi_{Z}$ of its shots in the first. The split-shot estimator is
unchanged, with $\hat f$ the sum of the two independent parts, so Theorem~1's
exact risk holds with the single-shot variance of each configuration replaced by
$V_{Z}/\pi_{Z}+V_{X}/\pi_{X}$. On a four-qubit model with $RY$ and $RZ$ layers and
CNOT rings ($p=16$, $B=16$), Monte Carlo (4,000 replicates, shots sampled from the
exact distributions) matched this risk at 16, 64, 256 and 1,024 shots per
configuration, both for $\pi_{Z}=1/2$ and for the split $\pi_{Z}\propto\sqrt{V_{Z}}$
that minimises the variance, within $1.0\%$ in every case (Monte Carlo standard
error about $0.3\%$). That split saved $7.5\%$ over equal halves, and the
exact-risk allocation across inputs, computed from the combined variance, saved
$3$ to $4\%$ over uniform on this model. Exploratory.

\section{Shot noise in a capacity measure}
\label{app:effdim}

\begin{figure}[htbp]
\centering
\includegraphics[width=\textwidth]{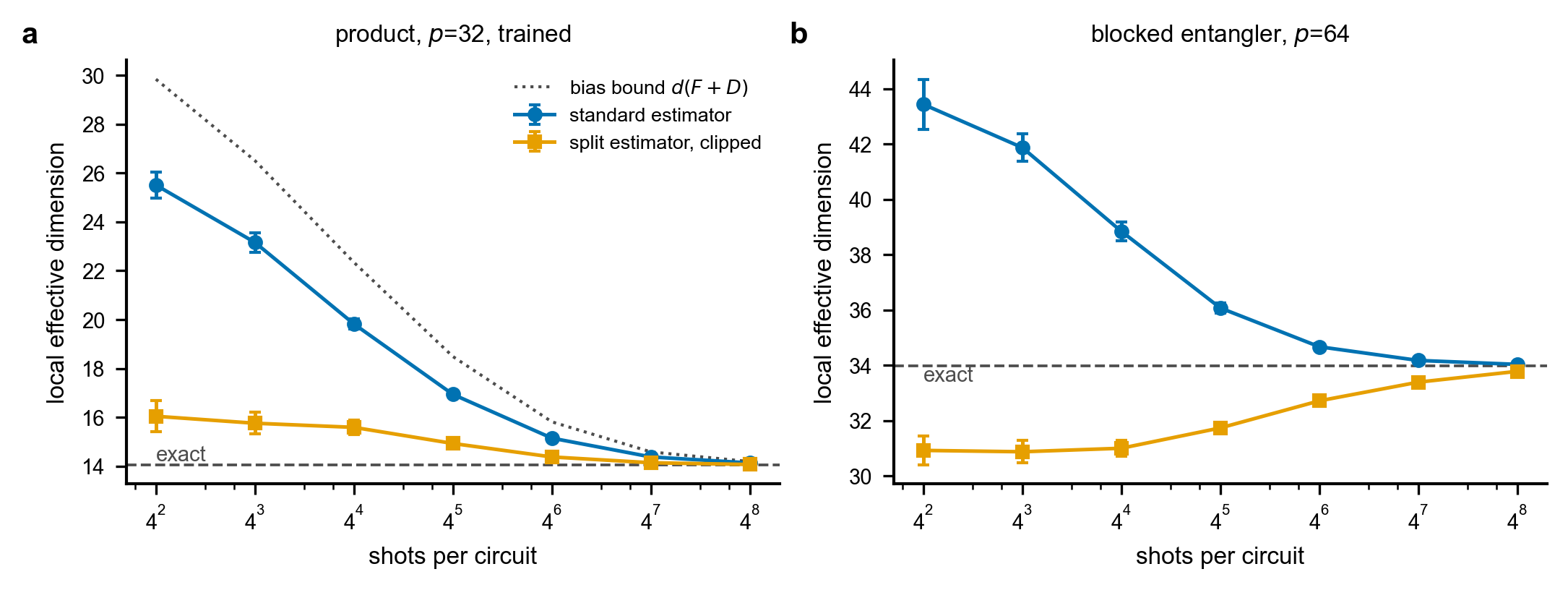}
\caption{\textbf{Shot noise inflates a capacity measure.} Local effective
dimension estimated from shots against shots per circuit, standard estimator (blue
circles) and split-shot estimator clipped at zero (orange squares), mean and
standard deviation over repetitions; the dashed line is the exact value and the
dotted line the bias prediction $d(F+D)$ (panel a). \textbf{a}~Trained product
model, $p=32$, $B=64$. \textbf{b}~Blocked circuit, $p=64$, $B=64$. Exploratory.}
\label{fig:effdim}
\end{figure}

The standard estimator $\hat F=B^{-1}\sum_{i}\hat g_{i}\hat g_{i}^{\mathsf T}$ has
mean $F+D(S)$, where $D\succeq0$ is the bias of Theorem~1. The main text defines
the local effective dimension $d=\sum_{m}\log(1+k\lambda_{m})/\log k$, with
$\lambda$ the spectrum of $pF/\operatorname{tr}F$ and $k=\gamma N/(2\pi\log N)$.
It is steepest where $\lambda$ is small: the bias fills near-null directions and
the model reads as more expressive than it is. The mechanism is the one by which
finite-shot features leave the ideal subspace in quantum linear
regression\cite{lomonaco2026}; here it acts on a capacity measure. On a trained
product model with $p=32$ and $B=64$ ($\gamma=1$, $N=10^{4}$) the exact local
effective dimension is $14.04$; measured with 256 shots per circuit it reads
$19.82$, and it comes within $0.5$ of the truth only at $16{,}384$ shots per
circuit, $6.8\times10^{7}$ executions (Fig.~\ref{fig:effdim}a). The split-shot
estimator, clipped at zero, gets there with a quarter of the shots on the same
grid. The bias prediction $d(F+D)$ bounds the measured value from above in every
cell and is within $19\%$ of it in every cell with $B>p$; at $B=p$ it overshoots
by up to $48\%$, and when $B<p$ the estimator's rank caps the inflation at $B$. On
the blocked circuit with $p=64$ and $B=64$ the exact value is $33.99$ and 256
shots per circuit read $38.85$ (Fig.~\ref{fig:effdim}b). There the clipped
split-shot estimator errs the other way, reading $31.01$, and converges no faster:
removing the bias of $\hat F$ does not remove the bias of a nonlinear function of
it.

Panels a to d of main-text Fig.~2 are an illustration of this mechanism on a
two-qubit model with one parameter per qubit and two training samples, computed
from exact gradients and simulated shots; they carry no result of their own.

\paragraph{From an error radius for $F$ to one for $d$.} Write
$\tilde F=pF/\operatorname{tr}F$, so that $d=\log\det(1+k\tilde F)/\log k$. To
first order in an error $\delta F$,
\begin{equation}
  \delta d\simeq\frac{k}{\log k}\operatorname{tr}\bigl[(1+k\tilde F)^{-1}\delta\tilde F\bigr],
  \qquad
  \delta\tilde F=\frac{p}{\operatorname{tr}F}\Bigl(\delta F-\frac{\operatorname{tr}\delta F}{p}\,\tilde F\Bigr),
  \label{eq:deltad}
\end{equation}
so $d$ depends on $\hat F$ only through $\hat F/\operatorname{tr}\hat F$, and the
error that controls it is the error of $F$ relative to $\operatorname{tr}F$. By
Cauchy--Schwarz,
$|\delta d|\le(k/\log k)\,\lVert(1+k\tilde F)^{-1}\rVert_{F}\,\lVert\delta\tilde
F\rVert_{F}$, and $\lVert\delta\tilde F\rVert_{F}\le
p\,(1+\sqrt{p/\mathrm{PR}(F)})\,\lVert\delta F\rVert_{F}/\operatorname{tr}F$. The
weight $(1+k\tilde F)^{-1}$ is close to 1 on the near-null directions of
$\tilde F$ and close to 0 on its large ones, which is why the bias $D(S)$, which
fills the near-null directions, moves $d$ most. The bound is loose by the factor
$k$; it shows which norm to aim for, and it does not say how many shots $d$
needs.

On a run with fewer inputs than parameters the inflation has no room to act. We
re-analysed a 78-qubit run of the companion paper\cite{companion} ($p=312$
parameters, $B=8$ inputs, 3{,}072 shots per circuit), drawing random subsets of
the shots. The standard estimate of $F$ has a Frobenius-relative error of $0.55$
at 24 shots per circuit and $0.12$ at 384, yet its local effective dimension,
with the same $k$, reads $13.63$ at 24 shots against $13.61$ for the noiseless
matrix. Both matrices have rank 8, so the shot-noise bias cannot enter the other
304 directions. Showing the inflation on hardware needs a run with more inputs
than parameters. Exploratory.

\paragraph{The inflation on a processor (registered).} The run that does so used
the trained product model above ($p=32$, $B=64$, exact value $14.04$) on
\device{ibm\_kingston}. The model has no two-qubit gates, so eight copies of its 16
qubits ran side by side on 128 qubits frozen before registration (the lowest
readout errors, median $0.8\%$, worst $2.1\%$), copy $c$ evaluating inputs
$i\equiv c \pmod 8$: 520 circuits of 4,096 shots, 568~s of quantum time, no
readout mitigation. Counts were kept per 32-shot chunk, which suffices for every
estimator here; estimates at fewer shots average 20 random subsets of chunks.
The standard estimator read $23.42$, $19.92$, $17.10$ and $15.29$ at 64, 256,
1,024 and 4,096 shots per circuit (standard deviations over subsets $0.33$,
$0.31$ and $0.15$), and the clipped split-shot estimator $15.76$, $15.74$, $14.93$
and $14.49$. Three registered checks, with thresholds fixed from simulation
before the run, passed. \emph{Inflation:} standard estimate at 256 shots over the
split-shot value at all shots $1.375$, threshold $1.20$ (simulation $1.38$).
\emph{Binomial size:} resampling binomial shots from the device's measured
frequencies gives $20.08$ at 256 shots; the difference is $0.011$ of the
reference, limit $0.05$, so drift and correlated readout add nothing visible.
\emph{Split-shot repair:} the split-shot excess at 256 shots is $0.23$ of the
standard one, limit $0.5$. Exact sampling of the design gave $1.38$, $0.006$ and
$0.22$, and 200 replicates under a readout-noise model gave no false failure.
The device's matrix at all shots is $1.03$ times the noiseless one in Frobenius
norm and lies $5.2\%$ from it, with participation ratio $2.10$ against $2.09$:
single-qubit circuits with low readout error barely damp it, so the inflation
measured here is the shot-noise effect of Fig.~\ref{fig:effdim}a, on a device.

\paragraph{An entangling model, and how well $d$ can be pinned down (registered).}
The second capacity run asked whether $d$ itself can be pinned down. The model is
the depth-2 CNOT brickwork on a line of 48 qubits ($RX(x)$ encoding with
$x\in\{0,\pi\}$, two layers of $RX(\theta)$ on every qubit followed by CNOTs on
even then odd pairs; $p=96$, $B=128$ random binary inputs, parity labels,
$f$ the mean $Z$). Its noiseless effective dimension is $40.78$, the
model-distribution Fisher matrix of Ref.~\cite{abbas2021b} gives $40.79$, and
the participation ratio is $15.9$. The gradients were measured with the
light-cone design of the companion paper\cite{companion}: 16 classes of 4 to 8
parameters with light cones apart, so each input needs 33 configurations. The best
96-qubit line of \device{ibm\_kingston} under the line rule of that paper
(readout error at most $0.10$, $T_{1}$ at least $20~\mu$s, every coupler below
$0.01$) was frozen before registration and cut into two lines of 48; input $i$ ran
on line $i \bmod 2$, two inputs per circuit. The arm used here ran every input's
33 configurations plus a drift copy of the unshifted circuit at 4,096 shots, the
$+$ and $-$ circuits of each class back to back. A second arm on inputs 0 to 7,
used in the companion paper, shared the jobs. In all, 3,328 circuits and
$1.05\times10^{7}$ executions in 10 jobs, dynamical decoupling, no twirling and no
readout mitigation, 2,772~s of quantum time. Counts were kept per 32-shot chunk.

Six registered checks passed; their thresholds were fixed from simulation, and
40 simulated runs of the whole design without bias failed the family of checks
once. Three concern this paper. \emph{Inflation:} the standard estimate at 256
shots, $65.96$, over the split-shot value at all shots, $45.30$, is $1.456$
(threshold $1.25$; simulation $1.51$). The curve reads $65.96$, $55.90$, $49.56$
(standard) and $48.73$, $46.30$, $45.32$ (split-shot) at 256, 1,024 and 4,096
shots. \emph{Bracket:} the Weyl bracket of Methods gives $[30.23,90.12]$,
half-width $29.9$ (limit 32), and the two halves of the run in time give point
values $45.66$ and $45.71$, each inside the other's bracket. \emph{Coverage:} the
block-minus-rest radius covered 16 of 16 blocks, and the exact risk predicted the
replicate error to a ratio of $0.991$. The other three, a test of the light-cone
assumption, its agreement with a direct bias test, and the saving of the light-cone
design, are reported in the companion paper; the direct test bounds the bias of the
light-cone readout at $0.169$ of a component's signal (one-sided 95\% upper value).

Secondary quantities. The device's matrix is $0.83$ times the noiseless one in
Frobenius norm and lies $20\%$ from it. The model-distribution effective dimension
from the same data is $45.26$. The split-shot point value sits $4.5$ above the
noiseless $40.78$ and $2.8$ above $42.47$, the mean split-shot reading in
simulation under the readout-noise model, whose true value is $40.84$;
how much of the remaining gap is gate noise and how much is the bias of the
clipped estimator for $d$ is not separable from one run. Rescaling $F$ by the
per-parameter ratio of joint to single shift differences that the companion
paper's direct test measures (median $0.90$) moves $d$ to $48.80$, an exploratory
sensitivity to the light-cone readout. The operator-norm radius is $0.39$ of the
mean eigenvalue. The half-width is not proportional to the radius, because the
lower end of the bracket sets every eigenvalue below the radius to zero: the run's
two halves in time, with radius $0.56$ of the mean eigenvalue, give half-widths of
$33.6$ and $33.5$ against $29.9$ at all shots. Evaluating the bracket with both
radii scaled as $S^{-1/2}$, $\pm0.5$ needs about $1.27\times10^{5}$ times the run's
shots.

\paragraph{An interval for $d$.} The split-shot value is stable but biased
($+1.46$, standard deviation $0.07$, over 40 simulated runs of the design under
the readout-noise model, true value $40.84$). Three corrections were compared on
the same runs. Richardson extrapolation in $1/S$, from the split-shot values at all
($d_{1}$), half ($d_{2}$) and a quarter ($d_{4}$) of the shots, $R_{2}=2d_{1}-d_{2}$
and $R_{3}=(8d_{1}-6d_{2}+d_{4})/3$, left biases of $0.52$ and $0.32$: the bias does
not fall as $1/S$ but about as $S^{-0.7}$. A bootstrap bias correction that
resamples chunks across the whole run over-corrects, to $-1.46$, because a
resampled chunk can fall in both halves and correlate them; resampling within
each half of a random partition keeps the halves independent and leaves $0.53$.
With $\pm1.96$ bootstrap standard deviations, these intervals covered the true
value in 0, 25 and 0 of 40 runs ($R_{2}$, $R_{3}$, half-preserving bootstrap).
Subtracting the mean residual bias of the other 39 runs as well, the intervals
covered in 40, 40 and 39 of 40. On the device data the corrected values are
$44.46$ \ci{44.14}{44.78}, $44.68$ \ci{44.25}{45.10} and $44.45$
\ci{44.20}{44.69}, so $d$ lies between $44.1$ and $45.1$ if the device's bias
follows the simulation's. That condition is not checked: on the device $R_{2}$ and
$R_{3}$ agree to $0.02$, against $0.20$ in simulation. Ledoit--Wolf shrinkage of
the 128 per-sample gradients moves $d$ to $77.2$ (shrinkage $0.14$): it removes
dispersion that belongs to $F$, not shot noise.

\paragraph{Corrections that need no noise model (exploratory).} Because the
corrected interval leans on the readout-noise model, we compared it with
corrections that use only the data, on the same 40 simulated runs per scenario
(true value about $40.8$) and on the device. The all-pairs U-statistic over the
shot chunks, which averages every way of splitting them, has bias $0.80$ against
$1.45$ for one random split; a delete-a-group jackknife of it (UJ) leaves $0.18$,
and the same jackknife of the split estimate leaves $0.35$. Their intervals are
too narrow for the bias they leave: UJ covered the true value in 27 of 40 runs, the
split jackknife in 13. The corrected estimate covered in 40 of 40 with bias $0.01$
under the model it was calibrated on. To test it under a model it was not
calibrated on, we damped every qubit's means by a fixed factor between $0.80$ and
$0.95$, and those of every shifted configuration by a further $0.96$, the
joint-over-single scale measured on the device, which stands in for gate noise
that the readout model lacks: the corrected estimate then covered in 36 of 40
runs with bias $0.20$, and UJ in 20 of 40 with bias $0.28$. A slow $4\%$ ramp of
every mean over the run changed neither. On the device the U-statistic
reads $44.80$, UJ $44.95$ \ci{44.72}{45.18}, the split jackknife $45.08$
\ci{44.80}{45.35} and the corrected estimate $44.45$ \ci{44.03}{44.88}; the
data-only corrections agree with the corrected one to within $0.5$, and the
capacity comparison does not depend on which is used ($12\%$ for UJ, $10$ to
$12\%$ for the corrected estimate). The plug-in $d$ with the data
count $N$ of the constant $\kappa=\gamma N/(2\pi\log N)$ set to $10^{3}$, $10^{4}$
or $10^{6}$ and $\gamma\in\{0.1,1\}$ inflates by $1.17$ to $1.48$ at 256 shots, so
the inflation is not an artefact of the constant.

\paragraph{A capacity comparison.} Classical tanh networks with 48 inputs and
one hidden layer of two units (101 parameters), on the run's 128 inputs and
labels with the same loss and the same matrix, exact, 200 random initialisations
with weights of variance $1/\text{fan-in}$: $d/p$ has 10th, 50th and 90th
percentiles $0.46$, $0.59$ and $0.69$. The quantum model's $d/p$ on the device is
$0.687$, $0.582$ and $0.516$ from the standard estimate at 256, 1,024 and 4,096
shots per circuit, $0.472$ from the split-shot estimate at all shots, $0.460$ to
$0.470$ from the corrected interval, and $0.425$ noiseless. The fraction of
networks it exceeds is $0.90$, $0.46$ and $0.24$ for the standard estimates,
$0.12$ for the split-shot value, $0.10$ to $0.12$ for the corrected interval and
$0.075$ noiseless. On small models the comparison is robust: for 4 and 6 qubits
with 24 to 48 parameters against networks with 13 to 33, $d/p$ is $0.72$ to
$0.89$ against $0.37$ to $0.45$, and no seed changed its verdict at 64 or more
shots per circuit (40 seeds per cell). In a choice between a shallow and a deep
ansatz on 6 qubits, where the two are close, the standard estimate changed the
verdict in $5$ to $8\%$ of seeds at 64 to 256 shots, and the clipped split-shot
estimate, biased low at so few shots, in $30\%$. Exploratory.

\paragraph{The comparison across parameter points (exploratory).} Is the reversal
on the 48-qubit run typical, or a property of one parameter point? We simulated the
depth-2 CNOT brickwork on a line of 16 qubits ($p=32$, $B=64$ binary inputs, parity
labels, mean-$Z$ readout), with coordinate-wise shifts read on all qubits. Shots are
Gaussian with the exact single-shot moments. The model was evaluated at 100 uniform
parameter draws and at 16 trained points (100 noiseless Adam steps from uniform
starts). The classical reference is the tanh network above with 37 parameters: 200
random initialisations for the uniform points, and the same 200 trained for 500
steps for the trained ones. At the uniform points the exact $d/p$ sits on average
at the 7th percentile of the classical distribution, and below its median at every
point. The standard estimate at 256 shots per circuit inflates $d$ by a factor of
$1.98$ on average. It places every one of the 100 points above the classical
median (mean percentile $99.98\%$); at 1,024 shots it still does at every point
(mean percentile $74\%$), and at 4,096 shots at 3 of 100 ($32\%$). The split-shot
estimate at 256 shots inflates $d$ by $1.24$ and places no point above the median
(mean percentile $19\%$); it moves the percentile by more than 20 points at 4 of
100. At the trained points the exact $d/p$ sits at the 65th percentile of trained
networks, above the median at 14 of 16. There the standard estimate at 256 shots
inflates $d$ by $2.9$ and places all 16 above the median (mean percentile $96\%$).
The split-shot estimate, unbiased for $F$ but not for $d$, inflates $d$ by $1.68$ at
these points and also places the two remaining points above the median; the
jackknife of the all-pairs U-statistic inflates it by $1.36$. The
parameter-averaged (global) effective dimension of Abbas et~al., computed from the
model Fisher matrices of the 100 uniform points, is $0.49$ exact and $0.82$ from the
standard estimate at 256 shots, close to the classical networks' $0.86$; the
split-shot estimate gives $0.55$. The forecast $d(F+D)$ exceeded the simulated mean
of the standard estimate at every point and shot count. The predicted excess was
$1.13$ to $1.28$ times the observed one (10th to 90th percentile over points), and
the rule $pV/(2GS)$ for $\operatorname{tr}D/\operatorname{tr}F$ agreed with the
exact ratio to within $0.3\%$ (median). Exploratory.

\paragraph{The comparison at six parameter points on hardware (registered).}
Table~\ref{tab:e132} gives the run's values. The three registered checks passed.
\emph{Inflation (passed).} The excess of the standard estimate at 256 shots over the
split-shot estimate at all shots was at least half the predicted excess at all six
points (five were needed). Observed over predicted ran from $0.70$ to $1.01$, lowest
at the trained points. \emph{Rank shift (passed).} The mean shift of the classical
percentile was $0.45$ against a threshold of $0.26$ (prediction $0.52$).
\emph{Forecast (passed).} The forecast $d(F+D)$ from the run's own chunk variances
was at least the observed standard reading minus two subset standard deviations
at all six points, and above the reading itself at all six, by $7$ to $9\%$. Two
things differed from the prediction. At the trained points the device's matrix has
an effective dimension two to three times the noiseless one: $15.6$, $13.9$ and
$9.7$ against $5.2$, $5.5$ and $4.4$. Readout errors do not produce this; the
readout model gives $5.3$, $5.5$ and $4.5$. A plausible cause, which we did not test, is
that the residuals are small at trained points, so gate noise that damps the
outputs changes the per-sample gradients most there. This is why the shift of the
classical rank at those points was 6 to 10 points against 14 to 23 predicted. The
second difference is a flaw of the plan. The registered bootstrap interval for the
split-shot estimate resamples chunks with replacement and then splits them into
halves, and it does not contain its own point estimate at any point. A chunk drawn
twice can land in both halves, which makes the halves dependent and biases the
product upwards. We therefore report the
standard deviation over halvings instead and do not use that interval. The run
was charged 2,328 quantum seconds.

\begin{table}[htbp]
\centering\footnotesize
\caption{\textbf{Six parameter points on \device{ibm\_marrakesh} (registered).}
Local effective dimension $d$ ($p=32$) from the standard estimate at 256 shots per
circuit (mean over 20 subsets), the split-shot estimate at 2,048 shots (mean over
20 halvings) and the jackknife (UJ, $\pm1.96$ se), the forecast at 256 shots, the
noiseless value, and the share of classical networks below the point.}
\label{tab:e132}
\begin{tabular}{lcccccc}
\toprule
Point & Standard, 256 & Split-shot & UJ & Forecast & Noiseless & Classical share, standard / split\\
\midrule
random 1 & $25.87$ & $14.61$ & $13.4\pm0.5$ & $27.6$ & $12.18$ & $100\%$ / $11\%$\\
random 2 & $26.51$ & $16.75$ & $16.8\pm0.7$ & $28.2$ & $14.58$ & $100\%$ / $22\%$\\
random 3 & $26.81$ & $16.24$ & $16.2\pm0.9$ & $28.6$ & $12.51$ & $100\%$ / $19\%$\\
trained 1 & $23.86$ & $15.58$ & $16.3\pm0.4$ & $25.7$ & $5.23$ & $100\%$ / $95\%$\\
trained 2 & $22.68$ & $13.91$ & $14.2\pm0.6$ & $24.5$ & $5.50$ & $100\%$ / $94\%$\\
trained 3 & $19.92$ & $9.66$ & $9.1\pm0.2$ & $21.8$ & $4.44$ & $98\%$ / $89\%$\\
\bottomrule
\end{tabular}
\end{table}

\section{Trained models on real data}
\label{app:trained}

Most simulated cells of the main text use independently drawn parameters and
parity labels. A registered experiment asks whether the law and the norm
dependence hold on trained models with real data, and whether training lowers the
effective rank.

\paragraph{Design.} Breast-cancer (569 samples, 30 features) and digits (0--4
against 5--9; 1797 samples, 64 features) from scikit-learn; 128 training and 256
held-out samples; standardised, projected by PCA, and angle-encoded as
$R_{X}(x_{q})$. Models: $R_{X}(\theta)$ layers with a brickwork entangler at
$n\in\{8,10,12\}$ and depth $D\in\{2,4,6\}$ (exact state vector), and the blocked
circuit at $n\in\{16,32,64\}$, $D=2$; readout the uniform mean of $Z_{q}$.
Parameters are drawn uniformly (\emph{random}) or trained from that draw by 60
full-batch Adam steps on the mean-square loss with exact adjoint gradients
(\emph{trained}; training loss falls to $0.77$ to $0.86$ of its initial value,
median by family and depth). The Fisher matrix is formed on the first
$B\in\{16,64,256\}$ held-out samples. Budgets at $\eps=0.15$ come from the exact
risk of the split-shot estimator in both norms; 144 rows.

\paragraph{Registered checks.} All three failed.
\emph{The minibatch through the effective rank.} The budget per unit effective
rank, $N_{F}/\mathrm{PR}$, rose from $B=16$ to 256 by a median factor $1.56$
(range $1.20$ to $1.84$), above the band $[0.67,1.5]$. At leading order the same
quantity changes by $1.02$, and $h$ and $V/G$ by less than $1\%$: the minibatch
enters the leading-order budget through the effective rank alone, as in the main
text. The rise is the finite-shot part of the exact risk. Averaging over more
samples lowers the shots each configuration needs (median $S$ falls to 53 at
$B=256$ in the trace norm), where the $1/S^{2}$ terms of Theorem~1 are no longer
small.
\emph{The effective rank as the price of the norm.} The ratio of the
Frobenius-relative to the trace-relative budget, over $\mathrm{PR}(F)$, has median
$0.83$ (range $0.58$ to $0.97$; band for the median $[0.9,1.1]$, every value inside
the registered $[0.5,2]$). At leading order the ratio is $1.000$ in every row, as
the identity requires. The trace-relative target needs fewer shots, so its exact
budget exceeds the leading order by more (median $46\%$ against $18\%$), and the
ratio falls with $B$: $0.92$, $0.82$ and $0.72$ at $B=16$, 64 and 256. The
effective rank is the price of the norm at leading order; at these budgets the
exact price is about $0.7$ to $0.9$ of it.
\emph{Training and the effective rank.} Training lowered the effective rank by a
median factor $0.90$ \ci{0.81}{1.05}, an interval that includes 1. We do not
claim that training lowers the effective rank.

\paragraph{Reported without a registered check.} The effective ranks of these
models are small, $1.2$ to $7.3$, and do not grow with the register on the blocked
family, so the fixed-cone exponent of the trained blocked models over $n=16$ to 64
is $2.08$ to $2.46$ in the Frobenius norm and $1.89$ to $2.08$ in the trace norm.
Training lowers the budget through the readout noise over the signal: $V/G$ falls
by a median factor $0.31$ (range $0.04$ to $0.77$), and the Frobenius budget by
$0.32$, while the effective rank does not move significantly. On random brickwork
parameters $V/G$ grows with depth (median $4.6$, $10$ and $23$ at $D=2$, 4 and 6),
as gradients concentrate; on the trained models it stays between $1.7$ and $3.1$.
The heterogeneity factor $h$ lies between $1.00$ and $1.68$, and is larger on the
trained models (median $1.09$ to $1.29$ against $1.02$ to $1.06$).

\begin{table}[htbp]
\centering
\footnotesize
\caption{\textbf{Trained and random models on real data} ($B=256$, $\eps=0.15$,
medians over $n$ and data set). $N_{F}$ is the Frobenius-relative execution
budget from the exact risk. Registered.}
\label{tab:trained}
\begin{tabular}{@{}lrrrrrrrr@{}}
\toprule
& \multicolumn{2}{c}{$\mathrm{PR}(F)$} & \multicolumn{2}{c}{$V/G$} &
\multicolumn{2}{c}{$h$} & \multicolumn{2}{c}{$N_{F}$}\\
Family, depth & random & trained & random & trained & random & trained & random & trained\\
\midrule
brickwork, 2 & 2.91 & 2.85 & 4.6 & 1.7 & 1.06 & 1.29 & $9.4\times10^{5}$ & $3.7\times10^{5}$\\
brickwork, 4 & 2.00 & 2.74 & 10.0 & 3.1 & 1.03 & 1.22 & $5.3\times10^{6}$ & $2.1\times10^{6}$\\
brickwork, 6 & 3.04 & 2.85 & 22.9 & 2.5 & 1.02 & 1.09 & $4.5\times10^{7}$ & $4.5\times10^{6}$\\
blocked, 2 & 4.00 & 3.28 & 2.4 & 1.3 & 1.06 & 1.21 & $5.9\times10^{6}$ & $2.6\times10^{6}$\\
\bottomrule
\end{tabular}
\end{table}

\section{Notation}
\label{app:notation}

Table~\ref{tab:notation} lists the symbols of the main text and these notes.

{\footnotesize
\setlength{\tabcolsep}{5pt}
\setlength{\LTcapwidth}{\textwidth}
\begin{longtable}{@{}>{\raggedright\arraybackslash}p{2.6cm}>{\raggedright\arraybackslash}p{11.4cm}@{}}
\caption{\textbf{Notation.}}
\label{tab:notation}\\
\toprule
\textbf{Symbol} & \textbf{Meaning}\\
\midrule
\endfirsthead
\toprule
\textbf{Symbol} & \textbf{Meaning}\\
\midrule
\endhead
\bottomrule
\endlastfoot
$n$ & number of qubits\\
$p$ & number of trainable parameters (gate angles) $\theta$\\
$B$ & minibatch size: number of training samples in the Fisher matrix\\
$x_{i}$, $y_{i}$ & input and label of sample $i$\\
$f$, $f_{i}$ & model output, the mean Pauli-$Z$ readout $n^{-1}\sum_{q}\langle Z_{q}\rangle$; its value on sample $i$\\
$r_{i}$ & residual $f_{i}-y_{i}$\\
$d_{i}$, $g_{i}$ & gradient of $f_{i}$; loss gradient $g_{i}=2r_{i}d_{i}$ of sample $i$\\
$F$, $\hat F$ & empirical Fisher matrix $B^{-1}\sum_{i}g_{i}g_{i}^{\mathsf T}$; its estimate\\
$\lVert\cdot\rVert_{F}$, $\operatorname{tr}$ & Frobenius norm; trace\\
$\eps$ & target relative precision (root-mean-square, relative to $\lVert F\rVert_{F}$ unless stated)\\
$S$ & shots (executions) per circuit configuration\\
$\BQ$ & total executions, $B(1+2p)S$ for the coordinate-wise estimator\\
$V$ & single-shot variance of $f$; $V_{c}$ at configuration $c$\\
$w_{j}$ & variance of the shift estimate of $\partial_{j}f$ per shot, about $V/2$\\
$v_{i}$ & single-shot variance at the unshifted configuration of sample $i$\\
$G$ & squared gradient norm $\lVert\nabla_{\theta}f\rVert^{2}$\\
$\kappa$ & alignment $\lVert F\rVert_{F}/\overline{\lVert g_{i}\rVert^{2}}$, with $\kappa^{2}=1/\mathrm{PR}(F)$\\
$\mathrm{PR}(F)$ & participation ratio of the spectrum of $F$, the effective rank\\
$\lambda_{k}$, $\lambda_{m}$ & eigenvalues of $F$; of $pF/\operatorname{tr}F$ in the effective dimension\\
$h$ & heterogeneity factor ($\ge1$) of residuals and signals across samples\\
$N^{\star}$ & fewest executions allowed by Theorem~2\\
$D$ & circuit depth in layers; also the bias matrix $D(S)$ of the standard estimator\\
$L$ & block length of the blocked circuit\\
$w$ & light-cone width (qubits a readout depends on)\\
$a$, $c$ & coefficients of the fit $\kappa^{2}\approx a/B+cL/n$\\
$d$ & local effective dimension\\
$k$, $\gamma$, $N$ & resolution $k=\gamma N/(2\pi\log N)$ of the effective dimension, with $N$ data points\\
$\mu_{c}$, $\rho_{c}$ & mean of $f$ and state at configuration $c$\\
$A_{c}$ & squared sensitivity $\lVert\partial F/\partial\mu_{c}\rVert_{F}^{2}$\\
$N_{c}$, $\nu_{c}$ & copies (executions) spent on configuration $c$; their fraction of $N$\\
$U_{r}$, $s$ & neighbourhood of the configuration states and its radius scale in the minimax risk\\
$\hat\rho$, $\delta$ & confidence radius; its nominal miss probability\\
\end{longtable}}

\section{Claim to evidence}
\label{app:claims}

{\scriptsize
\setlength{\tabcolsep}{4pt}
\setlength{\LTcapwidth}{\textwidth}
\begin{longtable}{@{}>{\raggedright\arraybackslash}p{4.2cm}>{\raggedright\arraybackslash}p{2.6cm}>{\raggedright\arraybackslash}p{3.4cm}>{\raggedright\arraybackslash}p{3.9cm}@{}}
\caption{\textbf{Claim to evidence.} Every numerical claim in the main text and
the supporting notes binds to a named field in a locked evidence record. Record
names are the file stems of the JSON archive described under Data availability;
records under \texttt{docs/followup} are exploratory scripts' outputs, and the
others are in \texttt{results}.}
\label{tab:claims}\\
\toprule
\textbf{Claim} & \textbf{Value} & \textbf{Record} & \textbf{Field}\\
\midrule
\endfirsthead
\toprule
\textbf{Claim} & \textbf{Value} & \textbf{Record} & \textbf{Field}\\
\midrule
\endhead
\bottomrule
\endlastfoot
$\kappa^{2}=1/\mathrm{PR}(F)$, every cell & $<10^{-9}$ & \rec{e104b_cone_exponent_vs_B} & \rec{identity_max_abs_dev}\\
$\kappa^{2}\approx a/B+cL/n$, 80 fixed-cone cells & $1.06$, $0.44$; $R^{2}$ $0.96$ (free constant $0.004$, $R^{2}$ $0.96$) & \rec{x1_kappa_refit} & \rec{fits.without_c0.coef}, \rec{.R2}; \rec{fits.with_c0}\\
Fixed-cone exponent, $n=64$--256, $B=12$ / $768$ & $2.12$ $(2.08,2.16)$ / $2.74$ $(2.69,2.79)$ & \rec{e104c_cone_exponent_vs_B_split} & \rec{fits.N_pred.F_n64to256.F_B*_n64to256}\\
Trace-relative exponent, $n=64$--256, $B=12$/48/192/768 & $1.95$, $1.94$, $1.92$, $1.85$; leading $1.96$ & \rec{e104c_both_norms} & \rec{table.*.trace_exact_rescaled_n64_256}\\
Slope of $\log\kappa^{2}$ in $\log p$ from the fit, $n=64$--256, $B=12$/48/192/768 & $-0.14$, $-0.39$, $-0.71$, $-0.90$ & \rec{x1_kappa_refit} & \rec{slope_log_kappa2_in_log_p_n64to256}\\
Rise $B=12\to768$, all $n$ & $0.47$ $(0.41,0.53)$ & \rec{e104c_cone_exponent_vs_B_split} & \rec{fits.N_pred.F_all_n.F_B768-F_B12}\\
Budget over $p^{2}V/(\eps^{2}\kappa^{2}G)$ & $2.03$--$3.77$ & \rec{e104c_cone_exponent_vs_B_split} & \rec{h_range_F}\\
Fixed-cone exponent gates, rise / exponent & $0.16$ $(0.10,0.23)$; $2.56$ $(2.51,2.61)$ & \rec{e104b_cone_exponent_vs_B} & \rec{fits.N_pred.F_all_n.F_B768-F_B12}, \rec{F_n64to256.F_B768_n64to256}\\
Split risk over Theorem~2 bound & $1.20$ to $1.0002$ & \rec{lower_bound} & \rec{C1.*.ratio}\\
Exact-risk allocation, gap to $N^{\star}$ & $3$--$8\%$ & \rec{lower_bound_alloc}, \rec{lower_bound_chain} & \rec{cells.*.bound_gap_exact}\\
Certificate coverage, product models & $0.935$--$0.965$ & \rec{lower_bound} & \rec{C4.*.coverage}\\
Allocation saving at scale & $1.61$--$4.44$ & \rec{e108_scale_checks} & \rec{cells.*.saving_pred}\\
Effective dimension, exact / 256 shots & $14.04$ / $19.82$ & \rec{effdim_cost} & \rec{cells.*.d_true}, \rec{rows.*.d_std_mean}\\
Effective dimension on device data, 24 shots / noiseless & $13.63$ / $13.61$ & \rec{x3_effdim_device} & \rec{grid.*.plugin_mean}, \rec{reference.d_noiseless}\\
Independent Monte Carlo of Theorem~1 & $|z|\le1.48$ & \rec{thm1_independent_mc} & \rec{max_abs_z}\\
Hardware, uniform job obs/pred & $1.09$ $(0.84,1.38)$ & \rec{e109b_hardware_replication} & \rec{gates.G1}\\
Hardware, allocation saving & $3.20$ $(2.42,4.17)$ vs $2.72$ & \rec{e109b_hardware_replication} & \rec{gates.G2}\\
Hardware, confidence-radius coverage & 62 of 64 & \rec{e109b_hardware_replication} & \rec{gates.G3}\\
Hardware, allocated job obs/pred & $0.85$ $(0.75,0.96)$ & \rec{e109b_hardware_replication} & \rec{jobB.ratio}, \rec{jobB.ratio_ci}\\
Hardware over noiseless efficient budget, registered / exploratory run & $1.11$ / $1.10$ & \rec{e109b_hardware_replication}, \rec{e109_hardware_certification} & \rec{multiplier_Nmin}\\
Hardware $\lVert F\rVert_{F}$ below noiseless, four jobs & $10$ to $18\%$ & same & \rec{job*.normF_pool}, \rec{exact_noiseless.normF}\\
$nV$, $nG$ seed means, blocked entangler, $n=16$ to 1024 & $0.88$--$0.90$, $0.38$--$0.41$ & \rec{kappa_saturation} & \rec{rows.*.nV}, \rec{rows.*.nG}\\
Hardware, exploratory run & $0.86$, $2.39$ vs $2.77$, 63/64 & \rec{e109_hardware_certification} & \rec{gates}\\
Hardware budget over noiseless & $1.10$ & \rec{e109_hardware_certification} & \rec{multiplier_Nmin}\\
Trained models: budget per PR, $B=16\to256$, exact / leading & $1.56$ / $1.02$ & \rec{e115_trained_models} & \rec{gates.G1}, \rec{rows.*.N_frob_lead}\\
Trained models: $(N_{F}/N_{\mathrm{tr}})/\mathrm{PR}$, exact / leading & $0.83$ / $1.000$ & \rec{e115_trained_models} & \rec{gates.G2}, \rec{rows.*.N_trace_lead}\\
Trained over random PR & $0.90$ $(0.81,1.05)$ & \rec{e115_trained_models} & \rec{gates.G3}\\
Trained over random $V/G$; $N_{F}$ & $0.31$; $0.32$ & \rec{e115_trained_models} & \rec{rows.*.V_over_G}, \rec{rows.*.N_frob}\\
Trained blocked exponent, $n=16$--64, F / T & $2.08$--$2.46$ / $1.89$--$2.08$ & \rec{e115_trained_models} & \rec{reported.blocked_trained_exponent_*}\\
Effective rank, trained and random, real data & $1.2$--$7.3$ & \rec{e115_trained_models} & \rec{reported.PR_range}\\
Effective dimension on hardware, standard estimate at 256 shots / split at all shots & $19.92$ / $14.49$ & \rec{e128_effdim_shot_bias_hardware} & \rec{analysis.plugin.256.mean}, \rec{analysis.reference_d_split_all_shots}\\
Effective dimension on hardware, bootstrap at 256 / split at 256 & $20.08$ / $15.74$ & \rec{e128_effdim_shot_bias_hardware} & \rec{analysis.boot.256.mean}, \rec{analysis.split.256.mean}\\
Entangling model on hardware, standard $d$ at 256 / split at all shots / inflation & $65.96$ / $45.30$ / $1.456$ & \rec{e129_certified_effdim_entangling} & \rec{analysis.d.curve.256.standard}, \rec{analysis.d.d_split_all}, \rec{analysis.d.inflation_256}\\
Entangling model, bracket for $d$; shot multiple for $\pm0.5$ & $[30.23,90.12]$; $1.27\times10^{5}$ & \rec{e129_certified_effdim_entangling} & \rec{analysis.d.bracket.d_low}, \rec{.d_up}, \rec{.shot_multiple_for_halfwidth_0p5}\\
Entangling model, model-distribution $d$ & $45.26$ & \rec{e129_certified_effdim_entangling} & \rec{analysis.d.d_model_fisher_split_all}\\
Bracket coverage in simulation, 40 runs & 40/40 & \rec{e129_d_bracket_calibration} & \rec{summary.coverage_d}\\
Entangling model, Frobenius radius, statistical / bias-aware & $0.047$ / $0.41$ & \rec{e129_certified_effdim_entangling} & \rec{analysis.certificate.radius_stat_rel}, \rec{.radius_bias_aware_rel}\\
Interval for $d$: coverage of $R_{3}\pm1.96$ sd; calibrated, union on hardware & 25/40; $[44.14,45.10]$ & \rec{x7_d_interval} & \rec{simulation.coverage.R3}, \rec{calibrated.union_of_hardware_intervals}\\
Ledoit--Wolf $d$ on hardware & $77.2$ & \rec{x7_d_interval} & \rec{hardware.ledoit_wolf.d}\\
Capacity comparison: share of classical networks below the quantum model, standard at 256 / corrected & $0.90$ / $0.10$--$0.12$ & \rec{x8_capacity_decision_flip} & \rec{hardware_decision.quantum_wins_fraction}\\
Capacity comparison: share below, standard at 1,024 / 4,096; split-shot at all shots & $0.46$ / $0.235$; $0.12$ & \rec{x8_capacity_decision_flip} & \rec{hardware_decision.quantum_wins_fraction}\\
Capacity comparison: jackknife; split-shot after the joint-to-single rescaling & $0.12$; $0.18$ & \rec{x8_capacity_decision_flip} & \rec{hardware_decision.model_free_UJ}, \rec{split_after_scale_correction}\\
Six points on hardware: standard at 256 / split-shot, random points & $25.9$--$26.8$ / $14.6$--$16.7$ & \rec{e132_capacity_across_draws_hardware} & \rec{analysis.points.*.standard}, \rec{split_all}\\
Six points: classical share, random points, standard / split-shot & $1.00$ / $0.11$--$0.22$ & \rec{e132_capacity_across_draws_hardware} & \rec{analysis.points.*.*.percentile}\\
Six points: registered checks & 3 of 3 pass & \rec{e132_capacity_across_draws_hardware} & \rec{gates}\\
Six points: forecast over standard reading & $1.07$--$1.09$ & \rec{e132_capacity_across_draws_hardware} & \rec{analysis.points.*.forecast_256}\\
Across 100 random points (simulation): share placed above the classical median, standard at 256 / split-shot / exact & $1.00$ / $0.00$ / $0.00$ & \rec{x14_capacity_across_draws} & \rec{summary.uniform.256}\\
Global effective dimension, exact / standard at 256 / classical & $0.49$ / $0.82$ / $0.86$ & \rec{x14_capacity_across_draws} & \rec{global_model_fisher_dnorm}\\
Corrections without a noise model: UJ on the device; coverage in simulation & $44.95$ $(44.72,45.18)$; 27 of 40 & \rec{x13_d_model_free} & \rec{hardware}, \rec{simulation}\\
Two-basis readout, exact risk against Monte Carlo, largest deviation & $1.0\%$ & \rec{x9_generality_checks} & \rec{two_basis_readout.max_abs_log_ratio}\\
Readout weights on hardware & $0.68$ $(0.48,0.98)$ & \rec{e113_readout_weighted_certification} & \rec{composition.weight_saving}\\
\midrule
One constant across families & CV $0.283$ & \rec{e46_readout_variance} & \rec{data.P3_parameter_free.ratio_cv_all}\\
Pooled certification constant & $0.344$ & \rec{e46_readout_variance} & \rec{data.P3_parameter_free.ratio_mean}\\
Product family to $n=256$ (measured) & $1.966$ $(1.934,1.997)$ & \rec{e45_large_n_scaling} & \rec{data.fit_logBQ.p_exponent}\\
Classes disjoint, no universal exponent & pass & \rec{e69_exponent_class_table} & \rec{T2_pass}, \rec{T3_pass}\\
Chain, bit-flip encoding, $B=64$ (measured) & $3.087$ $(2.645,3.529)$ & \rec{e42_estimator_family} & \rec{data.ps_exponent}, \rec{data.ps_exponent_ci}\\
Brickwork at scale (fixed-constant prediction) & $2.853\rightarrow1.715$ & \rec{e79_f1b_brickwork} & \rec{small_n}, \rec{scale.entangling}\\
Chain at scale (fixed-constant prediction) & $3.032\rightarrow1.758$ & \rec{e81_f1c_chain} & \rec{scale.entangling}, \rec{class_overlap_diagnostic}\\
$p_{\mathrm{eff}}$ saturates & true & \rec{e83_positional_attenuation} & \rec{predictions.P3_p_eff_flat_in_n}\\
No family reaches $nV$ growth & true & \rec{e84_readout_correlation_survey} & \rec{f1e_pass}\\
Fixed-cone exponent (fixed-constant prediction) & $1.984$ $(1.958,2.013)$ & \rec{e86_cone_slope_fe} & \rec{pooled_fe.A_fixed_w}\\
Proportional-cone exponent (fixed-constant prediction) & $3.034$ $(3.002,3.075)$ & \rec{e86_cone_slope_fe} & \rec{pooled_fe.B_prop_w}\\
Identity for fixed-constant predictions (algebraic), worst family & $0.0008$ & \rec{e86_cone_slope_fe} & recomputed from \rec{rows}; see note\\
Exact MSE vs Monte Carlo, 12 cells & $|z|\le1.29$ & \rec{e102_theorem1_checks} & \rec{exact_vs_mc.max_abs_z}\\
Leading order vs exact, $S=1024$ / $S\le64$ & $0.6\%$ / $26\%$ & \rec{e102_theorem1_checks} & \rec{leading_order_accuracy.*}\\
Split-shot estimator vs standard & $0.134$ vs $0.171$ & \rec{e102_theorem1_checks} & \rec{split_estimator.chain_B32_S32}\\
Neyman saving over uniform, max & $1.4\%$ & \rec{e102_theorem1_checks} & \rec{allocation.max_neyman_saving_product}\\
Theorem~1 on the product ladder & $1.965$ & \rec{e102_theorem1_checks} & \rec{e45_repredict.BQ_p_exponent}\\
Theorem~1 $\eps$-exponent on the ladder & $-1.99$ & \rec{e102_theorem1_checks} & \rec{e45_repredict.eps_exponent}\\
Cubic over-budget, $p=16\to512$ & $36$ & \rec{e102_theorem1_checks} & \rec{cubic_overbudget_factor_p16_to_p512.value}\\
$\BQ$ slope in $B$, product / chain & $0.50$ / $0.30$ & \rec{e101_minibatch_sweep} & \rec{e101.fits.*.BQ_slope}, \rec{BQ_slope_ci}\\
Chain shot slope vs prediction & $-0.70$ vs $-0.70$ & \rec{e101_minibatch_sweep} & \rec{e101.fits.chain.S_slope}, \rec{S_pred_slope}\\
Product: floor-censored cells; uncensored slope & $79/210$; $-0.79$ vs $-0.81$ & \rec{e102_theorem1_checks} & \rec{e101_diagnostics_nongating.product}\\
Constant at $B=4$ / $B=256$ (chain) & $0.54$ / $0.04$ & \rec{e102_theorem1_checks} & \rec{e101_diagnostics_nongating.chain.constant_at_B*}\\
Measured / predicted shots, minibatch sweep & $1.24$ & \rec{e101_minibatch_sweep} & \rec{e101.level.median_ratio}\\
Third-family constant (brickwork, $B=12$) & $0.341$ $(0.306,0.376)$ & \rec{e101_minibatch_sweep} & \rec{e101b.ratio_mean}, \rec{e101b.ratio_ci}\\
Blocked, fixed cone (measured) & $2.27$ $(2.20,2.33)$ & \rec{e103_measured_cone_window} & \rec{fits.B_Q.*.F}\\
Blocked, growing cone (measured) & $3.04$ $(2.72,3.32)$ & \rec{e103_measured_cone_window} & \rec{fits.B_Q.*.P}\\
Blocked, Theorem~1 prediction & $2.26$ / $3.02$ & \rec{e103_measured_cone_window} & \rec{fits.BQ_pred.estimate}\\
Blocked, fixed-constant prediction & $1.95$ / $3.04$ & \rec{e103_measured_cone_window} & \rec{fits.BQ_const.estimate}\\
Regime difference (measured) & $0.76$ $(0.45,1.01)$ & \rec{e103_measured_cone_window} & \rec{fits.B_Q.*.delta}\\
Measured / predicted shots at scale & $1.18$ & \rec{e103_measured_cone_window} & \rec{level.median_ratio}\\
Alignment $\kappa$, fixed cone, $n=16\to128$ & $0.46\to0.32$ & \rec{e102_theorem1_checks} & \rec{e103_kappa_mean_by_arm_n}\\
\end{longtable}}

\noindent\textbf{Note on the identity residual.} The record also stores
\rec{families.*.D2_predicted_exponent}, which is $2-\mathrm{d}\log(nG)/
\mathrm{d}\log n$ and omits the $nV$ term. The identity stated in the main text
carries both terms and is recomputed from \rec{rows} by the figure generator;
against the stored field the deviations are larger and are not the quantity the
paper reports.

\end{document}